\pdfoutput=1 
\documentclass[12pt]{article}        

\usepackage{graphicx}
\usepackage{amsmath}
\usepackage{siunitx}
\usepackage{booktabs}
\usepackage{url}

\usepackage{amssymb}
\usepackage{euscript}
\usepackage[toc,page]{appendix}

\usepackage{inputenc}

\usepackage{amsmath,amssymb,bbm,color}
\usepackage{amsbsy}
\usepackage{euscript}
\usepackage{graphicx}
\usepackage{hyperref}
\usepackage{cite}
\usepackage{enumerate}
\numberwithin{equation}{section}

\usepackage{alltt} 
\usepackage{color}

\newcommand{\flatDfDx}[2] {d\,{#1}/d\,{#2}}

\newcommand{\plus}   {{+}}
\newcommand{\minus}  {{-}}

\newcommand{\Wp}     {{W^+}}

\newcommand{\Wm}     {{W^-}}   

\newcommand{\pTW}    {{p_{T,W}}}

\newcommand{\Vckmsqr}[2]{{\left|V_{#1#2}\right|}^2}

\newcommand{\Asym}[1]   {\mathrm{Asym}^{(+,-)}\left(#1\right)}

\def\slashii#1{\setbox0=\hbox{$#1$}            
  \dimen0=\wd0                                 
  \setbox1=\hbox{\sl/} \dimen1=\wd1            
  \ifdim\dimen0>\dimen1                        
     \rlap{\hbox to \dimen0{\hfil\sl/\hfil}}   
     #1                                        
  \else                                        
     \rlap{\hbox to \dimen1{\hfil$#1$\hfil}}   
     \hbox{\sl/}                               
  \fi}

\begin{document}

\allowdisplaybreaks

\begin{titlepage}

\begin{center}
{\bf\Large High-luminosity Large Hadron Collider \\ with laser-cooled
isoscalar ion beams$\,{^\dag}$}
\end{center}
\vspace{0mm}

\begin{center}
{\bf M.~W.\ Krasny$\,^{a,b}$},
 {\bf A.~Petrenko$\,^{c,b}$} 
 {\em and\ } 
 {\bf W.~P{\l}aczek$\,^{d}$}
 \\
\vspace{1mm}

{\em $^a\,$LPNHE, Sorbonne Universit\'e, Universit\'e de Paris, CNRS/IN2P3, \\
                 Tour 33, RdC, 4, pl. Jussieu, 75005 Paris, France} \\
{\em $^b\,$CERN, Geneva, Switzerland} \\
{\em $^c\,$Budker Institute of Nuclear Physics, \\
Prospekt Akademika Lavrent'yeva~11, Novosibirsk, Russia} \\
 {\em $^d\,$Institute of Applied Computer Science, Jagiellonian University, \\
 ul.~{\L}ojasiewicza 11, 30-348 Krakow, Poland}\\

\end{center}

\vspace{0mm}
\begin{abstract}

\noindent
The existing CERN accelerator infrastructure is world unique
and its research capacity should be fully exploited. 
In the coming decade its principal {\it modus operandi}
will be focused on producing intense proton beams, accelerating and 
colliding them at the Large Hadron Collider (LHC) with the 
highest achievable luminosity.
This activity should, in our view, be
complemented  by new initiatives and their feasibility studies targeted on    
re-using the existing CERN accelerator complex in 
novel ways  that were not conceived when the machines were designed. 
They  should provide attractive, 
ready-to-implement research options for the forthcoming   
{\it paradigm-shift} phase of the CERN research.
This paper presents one of the case studies of the {\it Gamma Factory} 
initiative \cite{Krasny:2015ffb} -- a proposal of a new operation scheme of 
ion beams in the CERN accelerator complex.
Its  goal is to extend the scope and precision of the 
LHC-based research  by complementing the proton--proton collision programme 
with the {\it high-luminosity} nucleus--nucleus one.  
Its numerous physics highlights include studies of the exclusive 
Higgs-boson production in photon--photon collisions and precision
measurements of the electroweak (EW) parameters. 
There are two principal ways to increase  the LHC luminosity 
which do not require an upgrade of the CERN injectors:
(1) modification of the beam-collision optics and (2) reduction of 
the transverse emittance of the colliding beams. 
The former scheme is employed by the ongoing high-luminosity (HL-LHC) project.
The latter one, applicable only to ion beams, is proposed in this paper.
It is based on laser cooling of bunches of partially stripped ions 
at the SPS flat-top energy. For isoscalar calcium beams, 
which fulfil the present beam-operation constrains and which are particularly 
attractive for the EW physics,  the transverse beam 
emittance can be reduced by a factor of $5$ within the $8$\ seconds 
long cooling phase.
The predicted nucleon--nucleon luminosity 
of $L_{NN}= 4.2 \times 10^{34}\,$s$^{-1}$cm$^{-2}$
for collisions of the cooled calcium beams at the LHC top energy 
is comparable to the levelled luminosity for the 
HL-LHC proton--proton collisions, but with reduced pile-up background. 
The scheme proposed in this paper,
if confirmed by the future Gamma Factory proof-of-principle 
experiment, could be implemented 
at CERN with minor infrastructure investments.
\end{abstract}

\vspace{3mm}

\footnoterule
\noindent
{\footnotesize
\begin{itemize}
\item[${\dag}$]
This paper is dedicated to the memory of Evgeny Bessonov, 
the Gamma Factory group \\
member and our colleague, who passed away recently. 
\end{itemize}
}

\end{titlepage}

\newpage
\setcounter{tocdepth}{1} 
\setcounter{tocdepth}{2} 
\tableofcontents

\newpage

\section{Introduction}
\label{intro}

The Large Hadron Collider (LHC) is a  collider of partonic bunches   
containing a dynamical mixture of quarks, gluons and photons. 
The partonic bunch carriers which guarantee 
their stability over the beam acceleration 
and storage time are protons and stable nuclei (ions) 
characterised by their proton, $Z$,  and neutron,  
$A-Z$, content, where $A$ is the mass number of the nucleus. 

The use of the ion beams at the LHC has, so far, 
been largely confined to studies of the strong-interaction phenomena. The precision studies 
of the Standard Model (SM) electroweak (EW) interactions and searches 
for new, beyond the Standard Model (BSM),  
processes have been carried out using the proton beams. 
There were three obvious reasons for that. 

The first one was extending the collision energy of colliding partons to
its maximal value,  specified by the maximal magnetic field of the
LHC dipoles ---  partons carried by protons have their maximal
energies larger by at least a factor of two than
those carried by nuclei at the same parent-beam-particle magnetic
rigidity.

The second was minimising the multiplicity of
background particles obscuring the detection and measurement 
of the SM EW and BSM partonic processes. 
These background particles are produced in
ordinary, strong-interactions mediated, collisions of spectator
partons which accompany partons taking part in the SM EW
or BSM processes of interest. For the low luminosity LHC, 
characterised by a negligible multi-collision pile-up,  
the number of parasitic collisions of spectator partons
grows quickly with the nuclear mass number $A$ of colliding
particles. It is   minimal for the proton--proton ($pp$) collisions. 

The third and the most constraining reason was maximising
the partonic-collision luminosity to look for very 
rare BSM processes and reducing statistical errors of the
SM EW measurements. At the LHC, the number of partons 
transported to their
collision point in nuclear envelopes is significantly smaller
than the number of partons transported  by protons. 
This is due to several $Z$-dependent beam intensity-limiting 
effects, such as: the achievable yields of beam particles coming 
from the proton and ion sources, space-charge and intra-beam 
scattering effects in bunched beams, and -- for the large-$Z$ 
ions -- the presence of parasitic beam-burning processes 
affecting the beam safety in the superconducting rings. 

For the above three reasons the $pp$ collision scheme has always been  
assumed to be superior with respect to the nucleus--nucleus ($AA$) one.

In this paper we argue that -- in the forthcoming
high-luminosity (HL) phase of the LHC experimental programme -- the
above arguments in favour of  the use of the proton  
beams for the studies
of the SM EW and BSM phenomena lose their strength, or can be
circumvented by introducing a new operation scheme of ion  
beams which includes reduction of their transverse
emittance  by  laser-cooling.  

In order to find the optimal balance for the 
LHC research programs based on  
the proton and nuclear partonic bunches 
the following question have to be addressed:

\begin{itemize}
   
    \item
    What are the advantages of nuclei for which the  
    $u$ and $d$ valence-quark, photon and gluon
    composition can be tuned to their physics-goal-optimised 
    values?
    \item
    What is the optimal beam-particle choice which maximises 
    the rate of SM EW and BSM partonic collisions with respect 
    to the beam-burning parasitic collisions?
    \item
    What is the optimal beam-particle choice which minimises the 
    multi-collision pile-up background for the LHC bunched beams at the highest partonic luminosity?
    \item
    Which partonic bunch carrier maximises the effective rate of
    photon--photon collisions?
    \item
    Which beam particle allows for the  most precise  experimental
    control of the flavour-dependent fluxes and effective emittances of
    its quarks, antiquarks and gluons?
\end{itemize}

The analysis  presented in 
the first part of this paper leads to the 
conclusion that the isoscalar, $Z=A/2$, nuclei -- so far 
not considered as attractive partonic bunch  
carriers -- have numerous advantages with respect 
to protons in the high-luminosity phase of 
the LHC research programme, in particular for the 
precision measurements of the SM EW parameters 
and for the  detailed experimental 
investigation of the EW symmetry-breaking mechanism. 

From the historical perspective, such a conclusion is not original, 
as the progress in understanding the weak-interaction 
sector of SM in the 1970s, 1980s and 1990s
was  made by studying lepton scattering 
on isoscalar nuclei rather than hydrogen targets: 
(1) iron ($\mathrm{Fe}$)\footnote{$\mathrm{Fe}$
has a small excess of neutrons over protons, 
so the corresponding
non-isoscalarity correction had to be made.} 
in the CDHS experiment at CERN 
\cite{Berge:1989hr}, CCFR experiment at FNAL \cite{Auchincloss:1990tu}, 
NuTeV experiment at FNAL \cite{Tzanov:2005kr}, 
E140 experiment at SLAC \cite{Bosted:1988ts};
and (2) calcium ($\mathrm{Ca}$) in the CHARM experiment 
at CERN \cite{Allaby:1987bb}.
Special runs were proposed at HERA at DESY
with a deuterium beam \cite{HERA-Deuter}
to resolve the light-flavour structure of a proton. 
Finally, a fixed-target muon--deuterium scattering
experiment was  proposed at the SPS 
\cite{COMPAS-for-LHC}, as a support
experiment for the LHC precision-measurement programme. 
The goal of the latter two initiatives was to 
reduce the interpretation ambiguities
of the SM EW measurements at the LHC.

The most likely reason for which the beams of isoscalar 
ions have never been considered as portal to precision  
studies of the SM EW and BSM phenomena at the
LHC is that, so far, no scheme has been proposed  
to achieve high partonic luminosities in 
$AA$ collisions -- comparable to the ones for the proton beams. 

In this paper we propose such a scheme.
The underlying idea is to reorganise the electron-stripping 
sequence of the CERN ion
beams in order to allow ions to carry a small number of
attached electrons over their SPS acceleration cycle. 
The atomic degrees of freedom of the beam particles 
are used, in the proposed scheme, 
to cool the beam by laser photons at the top SPS energy, and  
reduce its transverse emittance. The beam-cooling phase,
lasting a couple of seconds, is then followed by stripping 
the remaining electrons in the SPS-to-LHC transfer line. 
The small-transverse-emittance fully stripped ion beam 
is then accelerated and brought to collisions with the 
counter-propagating beam in the LHC interaction points.

The isoscalar calcium beam is chosen for a concrete implementation 
of the proposed scheme. It satisfies the numerous beam 
operation constraints, discussed in this paper, 
and maximises its  physics potential
--  both in the EW and BSM sectors. 

Longitudinal laser cooling of low-energy and low-intensity atomic beams
in storage rings 
has already been demonstrated \cite{Schroder:1990zz,Hangst:1991zz}. 
The transverse laser cooling of such beams has been studied in \cite{Lauer:1998zz}.
The evaluation of various techniques of longitudinal and transverse 
cooling of atomic beams at ultra-relativistic energies will soon be  
addressed in the forthcoming R\&D phase of the Gamma Factory studies 
\cite{Krasny:2015ffb,Krasny:2018pqa} -- in its Proof-of-Principle (PoP) 
experiment \cite{GF-LoI}. If the Gamma Factory 
SPS PoP experiment confirms the simulation results presented 
in this paper and if the beam-cooling scheme is implemented, 
a path to high-luminosity operation of the LHC with nuclear beams 
will be wide open. Adding such a new LHC operation mode, as
discussed in this paper, could  extend  the scope and improve 
the precision reach of the LHC scientific programme. 
The proposed scheme could pave a new way to achieving the ultimate 
partonic luminosity at the future hadronic colliders, such as the FCC.

This paper is organised as follows.
In Section~\ref{Partonic_beams} we analyse the composition and 
properties of partonic bunches confined in protons and nuclei.
In Section~\ref{Merits} we discuss the relative merits of proton 
and nuclear envelopes of partonic bunches for the LHC physics programme. 
In Section~\ref{Luminosity} we identify dominant luminosity-limiting 
factors for the $AA$  collision scheme.
Laser cooling of ultra-relativistic atomic beams is discussed 
in Section~\ref{Cooling}. It is followed, in Section~\ref{Operation},
by the analysis of the beam operation constraints 
and the corresponding choice   
of the optimal beam particle species for their high-luminosity collisions
at the LHC.
Section~\ref{Ca-Ca} presents a proposal of a concrete implementation 
of the proposed scheme for the calcium beam.  
Finally, Section~\ref{conclusions} contains conclusions and outlook. 
In Appendix~\ref{annex:kinematics} we discuss the technical aspects 
of kinematics and dynamics of photon absorption and emission by
highly ionised atoms, relevant for this paper.

\section{Protons and nuclei as carriers of partonic bunches}
\label{Partonic_beams}

\subsection{Space-time picture}
\label{Compostion}

Protons and nuclei colliding at the LHC, when ``observed" in the
rest-frame of the  counter-propagating beam with high space-time
resolution, can be considered as bunches of quasi-independent, 
point-like partons: quarks, gluons and photons. The
principal difference of the bunches of these virtual partons and the
classical bunches of protons is that the intra-bunch
dynamics of the former  is driven solely by the 
partonic-bunch-internal 
QCD and QED interactions. These interactions are by several  orders 
of magnitude stronger than the intra-beam and extra-beam interactions 
of proton bunches.  As a consequence, the emittance
of partonic bunches is independent of the LHC-ring lattice and constant over 
the acceleration and storage cycle. 

Each parton is confined within a fraction of the partonic-bunch 
space-time volume. This volume is fully determined by the energy, 
$E_p$, of the counter-propagating parton probing the inner 
structure of the partonic bunch, its 
energy-loss, $\nu$, and momentum-loss, $q$, in the space-time volume 
occupied by the parton. 
The principal difference of the proton 
and nuclear partonic bunches reflects their specific 
valence-quark, sea quark, gluon and photon composition as well as
the $A$- and $Z$-dependent momentum distributions of their 
components. 

The partonic distributions in the proton and nuclear
bunches can be related to each other by simple formulae 
discussed in the subsequent sections.
An extended discussion presenting the present status of the phenomenology  
of partonic distributions is presented in
Appendix~\ref{annex:PDFa}.

\subsection{Quarks}
\label{Quarks}

The flavour  of partonic bunches is determined by 
the atomic  $Z$ and mass $A$ numbers of their carriers.
Protons contain two valence $u$ quarks and one valence $d$ quark. 
Nuclei contain  $A+Z$ valence $u$ quarks and $2A-Z$ 
valence $d$ quarks. While the flavour composition of partonic 
bunches plays a very limited  role in the strong interaction processes, 
it plays a central role in electromagnetic and weak processes. 
This is because $u$ and $d$ quarks carry different electric 
charges, $Q^{u,d}$,  different weak isospin $t_{3L}^{u,d}$ 
and  have different axial and vector couplings, 
$g_V^{u,d} = t_{3L}^{u,d} + 2Q^{u,d}\sin^2\theta_W$ and $g_A^{u,d} = t_{3L}^{u,d}$,
to the EW $Z$-boson. 
As a consequence, the choice of the beam particle determines  
the beam-type-specific characteristics of collisions involving 
production of lepton pairs, $W$, $Z$ and Higgs bosons.

The valence-quarks, together with their associated quark--antiquark 
sea pairs (of various flavours: up, down, strange, charm, bottom) 
and gluons, form -- within the nuclear-bunch volume,
defined by its radius $R_A  = 1.2 \times A^{1/3}\,$fm -- clusters
containing two valence $u$ quarks and one valence $d$ quark 
(virtual protons) and two valence $d$ quarks and one valence 
$u$ quark (virtual neutrons). 

Since the binding energy of these virtual nucleons is 
significantly  smaller than the nucleus mass, 
the distributions 
$ u_{v}^{A,Z}(x_A,k_t,Q^2)$ and $d_{v}^{A,Z}(x_A,k_t,Q^2)$
of the fraction $x_A = p_q/p_A$ of the 
nuclear-partonic-bunch momentum carried 
by a valence-quark and of its  transverse momentum $k_t$
can be expressed at an arbitrary resolution scale $Q^2 = \nu ^2 - q^2$ 
in terms of  the corresponding distributions,  
$u_{v}^p(x_p,k_t,Q^2)$ and $d_{v}^p(x_p,k_t,Q^2)$,  for unbound protons:
\begin{equation}
     u_{v}^{A,Z}(x_A,k_t,Q^2) \approx (A-Z)d_{v}^p(x,k_t,Q^2) + Zu_{v}^p(x,k_t,Q^2)  \\
     \label{eq:1}
\end{equation}
and
\begin{equation}
      d_{v}^{A,Z}(x_A,k_t,Q^2) \approx (A-Z)u_{v}^p(x,k_t,Q^2) + Zd_{v}^p(x,k_t,Q^2) 
      \label{eq:2}
\end{equation}
by setting $x_p = p_q/p_p = A x_A = x$.

The corresponding momentum distributions of the $u$, $d$, $s$, $c$ and $b$ 
sea quarks and antiquarks, $q_s$, are also well approximated by
\begin{equation}
      q_{s}^{A,Z}(x_A,k_t,Q^2) \approx  Aq_{s}^p(x,k_t,Q^2). 
      \label{eq:3}
\end{equation}
Note that the isospin symmetry of the strong interactions implying: $u^p = d^n$ and $d^p = u^n$ 
has explicitly been used in the above formulae.   

In reality, quark clusters are confined within the space-time volumes 
which are slightly larger than the volume of the free protons and neutrons
and are surrounded by the nuclear-density-dependent virtual pion/kaon cloud. 
This meson cloud binds quark clusters within the nucleus volume. 
Finally, the quark clusters move within the nuclear bunch volume 
with the Fermi-motion velocities.

To account for the above nucleon binding effects, the quark and antiquark distributions 
in free protons, $q_{v,s}^p(x,k_t^2)$, have to be replaced in Eqs.~(\ref{eq:1}), 
(\ref{eq:2}) and (\ref{eq:3}) by the modified distributions (see e.g.\ \cite{Eskola:2017rmp}):
\begin{equation}
      q_{v,s}^{p/A}(x,k_t,Q^2) = R^A_{v,s}(x,k_t,Q^2) \times q_{v,s}^p(x,k_t,Q^2). 
\label{eq:RAq}
\end{equation}

\begin{figure}[!htpb]\centering
	\includegraphics[width=1.0\linewidth]{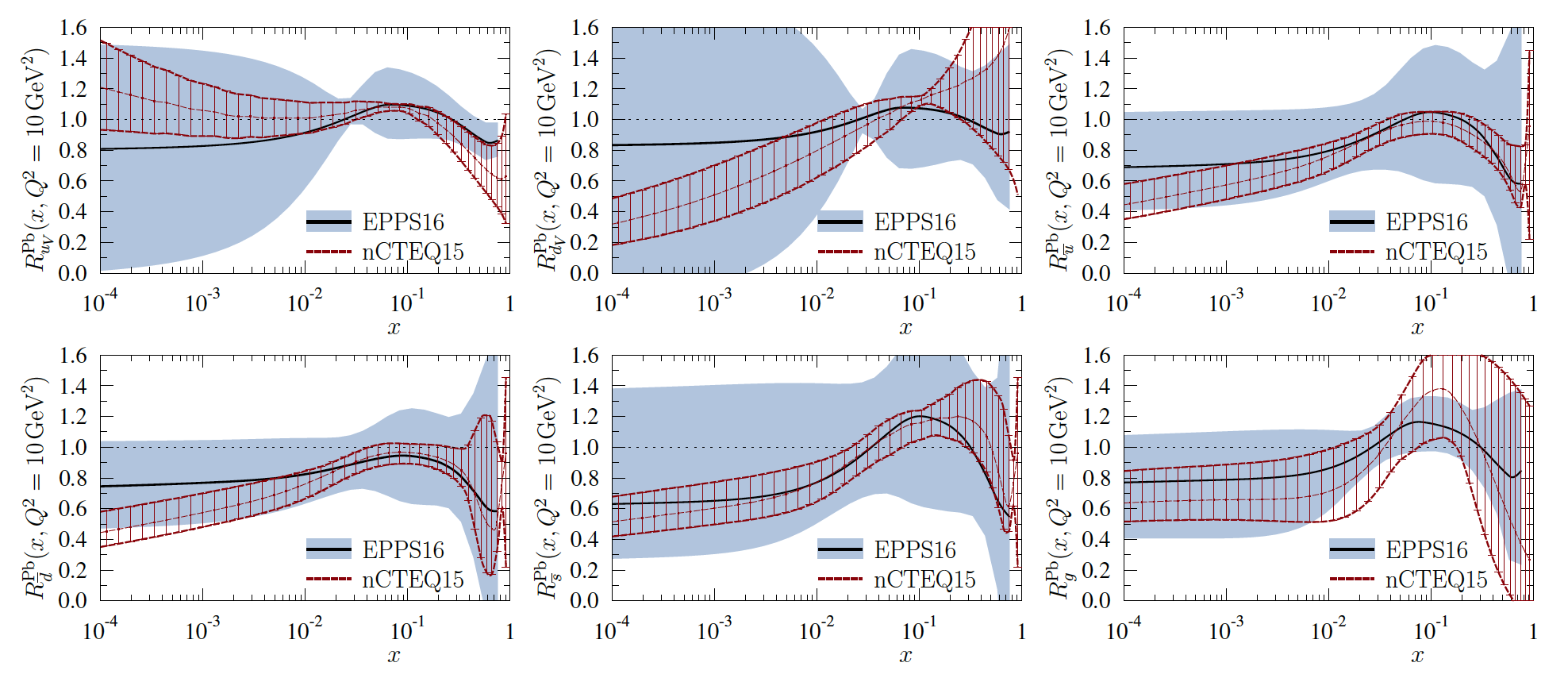}
	\caption{Nuclear modification factors for proton distribution functions 
	of quarks and gluons, taken from  \cite{Eskola:2017rmp}.}
    \label{nucl_modifications}
\end{figure}

The correction factors $R^A_{v,s}$ describe those of 
the strong interaction effects which cannot be 
controlled by the present perturbative computational methods 
of the theory of the strong interactions, the quantum chromodynamics (QCD), 
and must be measured. 
They quantify the 
approximations present in Eqs.~(\ref{eq:1}), 
(\ref{eq:2}) and (\ref{eq:3}).
It is important to note that the presence of the  non-perturbative QCD  
correction factors does not modify the QCD evolution 
equations \cite{DGLAP}, providing 
a relationship of partonic distribution at a fixed  and large 
resolution scale $Q^2$,
except a very low $x$ region (irrelevant to this paper).
In Fig.~\ref{nucl_modifications} we recall their size, integrated 
over the transverse momentum of the partons, $k_t$, as extracted from 
the experimental data  by the Eskola {\it et al.}~\cite{Eskola:2017rmp} and 
CTEQ~\cite{Kovarik:2015cma} groups. The correction factors 
do not differ significantly from $1$ over 
the full range of the $x$-variable.

\subsection{Photons}
\label{photons}

Nuclear partonic bunches carry a sizeable number of photons.  
Their  photon content grows very fast with increasing charge 
of the nucleus, proportionally to $Z^2$. 
For the largest-$Z$ nuclei,  photons are by a factor of $\sim 10^4$  
more abundant than for the equivalent-energy  protons. 
This property,  specific to QED and absent in QCD, is 
confined to the kinematical regime in which the wavelength of 
the photon is larger than the nuclear bunch size. Such photons 
do not resolve the internal partonic-bunch structure and originate  
from  a coherent action of all the charged constituents of partonic bunches.  
The coherence condition is fulfilled up to a 
maximum energy of the quasi-real photons of 
$ \omega \leq \gamma_L/R_A$, 
where $\gamma_L$ is the Lorentz factor of the beam particle and $R_A$ is the radius of the nucleus.
The transverse momentum of these photons satisfies the condition:
$k_t \leq 1/R_A$. 
At higher photon energies, up to the maximum 
energy of $ \omega \leq \gamma_L/R_N$, where $R_N$
is the nucleon radius, the photon content of nuclear bunches is 
expected to rise  proportionally to the  beam-particle charge number $Z$. 
Finally, at still higher energies 
the photon content is  expected to rise with increasing size 
of the nucleus proportionally to $A +Z/4$.

\subsection{Gluons}
\label{gluons}

Gluons are the most abundant constituents of the proton 
and nuclear bunches. Because of the colour confinement, gluons
cannot propagate freely and are confined to the distances 
which are smaller than the  nucleon diameter. 
Nuclear partonic bunch is a medium which does not conduct 
the colour current at the distances exceeding the nucleon size.
As a consequence, the gluon content of the nuclear partonic bunches, 
contrary to the photon content, is not enhanced by the colour-charge 
coherence effects. It rises significantly slower with increasing 
size of the nucleus than the photon content.

The gluon momentum distributions for the nuclear bunches can 
be expressed in a similar way as the  
sea-quark distributions by the following expression: 
\begin{equation}
 G^{A}(x_A,k_t,Q^2) = A \times  R^A_G (x,k_t, Q^2) \times G^p(x,k_t,Q^2).
\label{eq:RAg}
\end{equation}
It is important to note that  since QCD  
is  flavour-invariant, 
the gluon content of nuclear bunches rises proportionally to $A$ 
and is independent of the electric charge $Z$ of the nucleus.

In Fig.~\ref{nucl_modifications} the gluon nuclear modification 
factor $R^A_G$  is presented in the lower-right corner. 
As expected, it is close to $1$ and has a similar shape as  
the corresponding nuclear modification factors for the sea quarks.

\section{Merits of proton and nuclear partonic bunches}
\label{Merits}

\subsection{High-energy frontier of partonic collisions}
\label{energy_frontier}

The equivalence of the fractional momenta distributions of 
the quarks and gluons 
in proton and nuclear bunches can be translated 
to the equivalence of their momentum distributions 
only if the following relation between the total momenta 
of the proton and nuclear bunches is fulfilled:  $p_A = A \times p_p$.
In reality,  for the  fixed beam 
magnetic rigidity, the maximal beam momentum of the nuclear 
beam is always lower by at least a factor of two
than that of the proton beam: $p_A = (A \times p_p)/Z$.

There are two consequences of this constraint. They 
clearly demonstrate the unique merits of the  proton
partonic bunches at the LHC.
Firstly, the high-energy frontier of partonic
collisions, specified by the condition $x >0.5$, is not accessible
for  collisions of nuclear bunches. Secondly, since the 
partonic momentum distribution decreases with increasing $x$, there 
will always be a penalty of smaller partonic luminosities for collisions
of bound w.r.t.\ free nucleons (protons).  The ``penalty factors'', 
defined as the ratio of partonic fluxes 
for the nuclear and proton bunches at the same beam magnetic 
rigidity,  are sizeable in the region of $0.2 \leq x \leq 0.5$, 
in particular for collisions of gluons and sea 
quarks. They are small in the $x \approx 0.01 $ region -- the domain of interest for 
production of the $W$, $Z$ and Higgs bosons at the LHC: 
in the range of  $1.1$--$1.3$ for 
$u$ and $d$ quarks, and in the range of $1.5$--$1.7$ for the 
$s$, $c$, $b$ quarks and gluons, at the resolution scale $Q^2 = M_Z ^2$. 

\subsection{Precision frontier of partonic collisions} 
\label{Precision}

The interpretation of the measurements at the LHC requires a precise
knowledge of the content and momentum distributions of partons 
in beam particles.

Proton partonic bunches appear to have an important advantage over the nuclear ones because the momentum distributions of partons in proton are, at present, known to higher precision. 
This is, to a large extent, related to the fact that the HERA 
collider -- one of the principal sources of the information on 
partonic distributions -- was running only in the electron--proton 
collision  mode. The electron--nucleus collision mode was proposed 
in \cite{DESY_workshop,HERA-eA}. It was  developed initially 
as the  DESY-internal project 
\cite{Arneodo:1996qa,Krasny:1997mv}, and subsequently 
as a joint DESY--GSI
project \cite{HERA-eA-GSI}. Following the DESY decision to choose, 
as the future HERA operation, mode the ``high luminosity"
electron--proton rather than electron--ion collisions, 
the project migrated to BNL 
and became the eRHIC project. 
The BNL-based electron--ion collider option was presented, 
for the first time, 
in \cite{Krasny:2001nin}. Its  physics and  detector concepts, 
together with the accelerator requirements were developed further in 
\cite{Krasny:1999av,Snowmass_physics,Snowmass_detector, Krasny:2001tr}.
The first eRHIC white paper was published  in 2002 \cite{eRHIC}. 
The project had to wait until the completion of the FRIB
\cite{Bollen:2010leu} construction to become, 
following extensive accelerator design studies both at BNL and Jefferson Laboratory \cite{Accardi:2012qut}, the leading 
future accelerator-infrastructure project in 
the USA. One of its highlights is the 
precise measurement of the nuclear partonic distributions. 

The present advantage of the proton bunches fades away
in the high-luminosity, precision phase of 
the LHC experimental programme -- in
particular for the basic EW measurements, 
such as e.g.\ the measurements of the $W$-boson mass and 
$\sin^2\theta_W$. As argued  in \cite{Krasny:2010vd}, the 
LHC restricted to its $pp$ collision mode cannot
improve their measurement precision that has already been achieved 
at the previous particle colliders.
This is caused both by the limited  statistics of 
the HERA charge--current as well as charm and beauty production 
events\footnote{The proposed  HERA upgrade 
programme included rebuilding of the injectors: 
Linac and DESY3. This investment was necessary to increase  
the HERA luminosity to a level that could have  satisfied the LHC 
precision-programme requirements.},
and by the insufficient flavour-dependent constraints of 
partonic distributions coming from the analysis of 
the EW bosons and Drell--Yan lepton-pair production 
processes at the LHC\footnote{These processes do not suffer 
from uncertainties caused by an insufficient theoretical 
control of the hard-processes matrix elements and final-state 
interactions of  quarks and gluons.}. 
While the precision of the LHC constraints will certainly 
improve with the increased statistics and development of 
more precise theoretical frameworks of the data analysis, 
no new experiments are planned to improve the precision 
of the necessary  LHC-external constraints.  Consequently, 
no matter what effort is made at the LHC, the precision 
of the several LHC key measurements 
will always be limited by the LHC-external data -- 
as long as the proton--proton collisions remain 
the exclusive high-luminosity mode of the collider operation.

\subsection{Isospin symmetry}
\label{Symmetries}

The way to achieve the full, {\em in situ}, experimental control 
of the momentum distributions of partonic bunches at the LHC, proposed
in this paper, is to exploit symmetries in the interactions of their constituents.

The proton bunches contain a mixture of $u$ and $d$ quarks which 
carry different electric charges $Q^{u,d}$ and  different 
weak-isospin $t_{3L}^{u,d}$ components. 
Such bunches, as far as the EW processes are concerned,
are equivalent to bunches containing a mixture of electrons and  
electron-neutrinos with their relative proportion known 
to a limited accuracy. 

There is, however,  an important difference in the above two cases.
The pure electron and neutrino beams can be produced, 
while the pure beams of $u$ and $d$ quarks cannot.
The solution proposed below is to choose their optimal and 
precisely controlled mixture. 
Such an optimal mixture is provided by the isoscalar, $Z=A/2$, nuclei
for which -- thanks to the isospin symmetry of the strong 
interactions -- the momentum distributions of the $u$ and $d$ quarks are 
the same\footnote{A tiny violation of the equality may become 
important for ultra-high-precision measurements. The 
corresponding corrections, reflecting different electric 
charges of the $u$ and $d$ quarks,  are however theoretically 
well controlled by QED.}:
\begin{equation}
     u_{v,s}^{A=2Z,Z}(x,k_t,Q^2) =  d_{v,s}^{A=2Z,Z}(x,k_t,Q^2).
     \label{iso-constrains}
\end{equation}
Isoscalar-nuclei contain, like protons, a mixture of $u$ 
and $d$ quarks. What differs them from protons is that 
partonic momentum distributions, both in the valence and in 
the sea sectors and for all the $Q^2$  scales, become 
interrelated by the above symmetry relations. 

Restoring the isospin symmetry of the first generation quarks 
can play a similar role  in reduction of the LHC data 
interpretation ambiguities as exploiting the matter--antimatter 
symmetry for left--right symmetric detectors at the Tevatron collider. 
As we shall discuss in the next section, the isospin symmetry 
of isocalar nuclei allows to fully constrain the  momentum 
distributions of the partonic bunches solely by the LHC data.

\subsection{Flavour structure of isoscalar 
             partonic bunches}
\label{Constraints}


At the LHC energies, the number of independent quark and antiquark 
momentum distribution functions describing proton and nuclear 
bunches (for five quark flavours: $u$, $d$, $s$, $c$ and $b$) is ten. 
The equality of the sea quark and antiquark distributions of 
the s, c, and b flavours, $s=\bar s$,  $c=\bar c$ and $b=\bar b$,  
is a consequence of the gluonic-excitation  
origin of the heavy-flavour component of partonic 
bunches\footnote{For the $s$ quarks, a small violation of this equality
is likely and may call for a small correction to be applied;  
note that for the u and d quarks at $x \sim 6 \times 10^{-3}$ 
such an equality is violated at the level of 
$\sim$15\% for protons.}. The above three constrains reduce
the number of needed flavour-dependent distributions from ten to seven.

The measurements of momentum distributions of 
charged leptons in the $Z$, $W^+$, $W^-$ and 
non-resonant lepton-pair production processes 
constrain five out 
of seven unknown quark flavour-dependent distributions\footnote{In the next-leading-order (NLO) QCD framework, the gluon momentum 
distribution influences as well the measured momentum distributions
of produced leptons. This distribution is, however,  constrained by the 
QCD evolution equations \cite{DGLAP} which control the relationship between 
the quark (antiquark) and gluon distributions.
The relation of the partonic distribution at the two 
resolution scales: $Q^2 = M_Z^2$ and $Q^2 = M_W^2$ is 
constrained by the dedicated measurement procedure 
discussed in detail in \cite{Krasny:2010vd}.}, 
leaving the remaining 
two to be unconstrained. Note that for leptons produced 
at the $Z$-resonance peak, the distributions of both the 
negatively and positively charged lepton can be used. 

For the proton bunches, the missing two constraints must  
be provided by the LHC-external data. The flavour content 
of  the isoscalar partonic   
bunches is, on the contrary, fully constrained by the
two isospin-symmetry relations (for the valence 
and sea quarks) expressed  by the formula~(\ref{iso-constrains}). 
Therefore, for the beams of isoscalar nuclei, the precision 
of the LHC measurements does not any  longer depend upon  
the LHC-external constraints.  Moreover, for the isoscalar nuclei,  
special observables can be defined
\cite{Krasny:2010vd}, which facilitate the determination 
of specific partonic 
distributions. This is illustrated in the following example, 
discussed in detail in \cite{Fayette:2008wt}.

Let us consider the $W$-boson charge asymmetry observable,
$\mathrm{Asym}^{(+,-)}(\pTW)$ defined as: 
\begin{equation}
    \Asym{a} \; =\; 
    \frac
        { \flatDfDx{\sigma^\plus}{\pTW} - \flatDfDx{\sigma^\minus}{\pTW} }
        { \flatDfDx{\sigma^\plus}{\pTW} + \flatDfDx{\sigma^\minus}{\pTW} },
        \label{eq_def_charge_asym}
\end{equation}
where $\plus$ and $\minus$ refer to the electric charge of 
the $W$~boson, $\flatDfDx{\sigma^\pm}{\pTW}$ is the differential 
cross section and $\pTW$ is the transverse momentum of the $W$-boson.
The  asymmetry of the $\pTW$ distribution reflects
the flavour asymmetries in the distributions of transverse 
momentum $k_t$ of quarks and antiquarks producing 
the $\Wp$ and $\Wm$ bosons. 

\begin{figure}[htpb]\centering
    \includegraphics[width=0.49\linewidth]{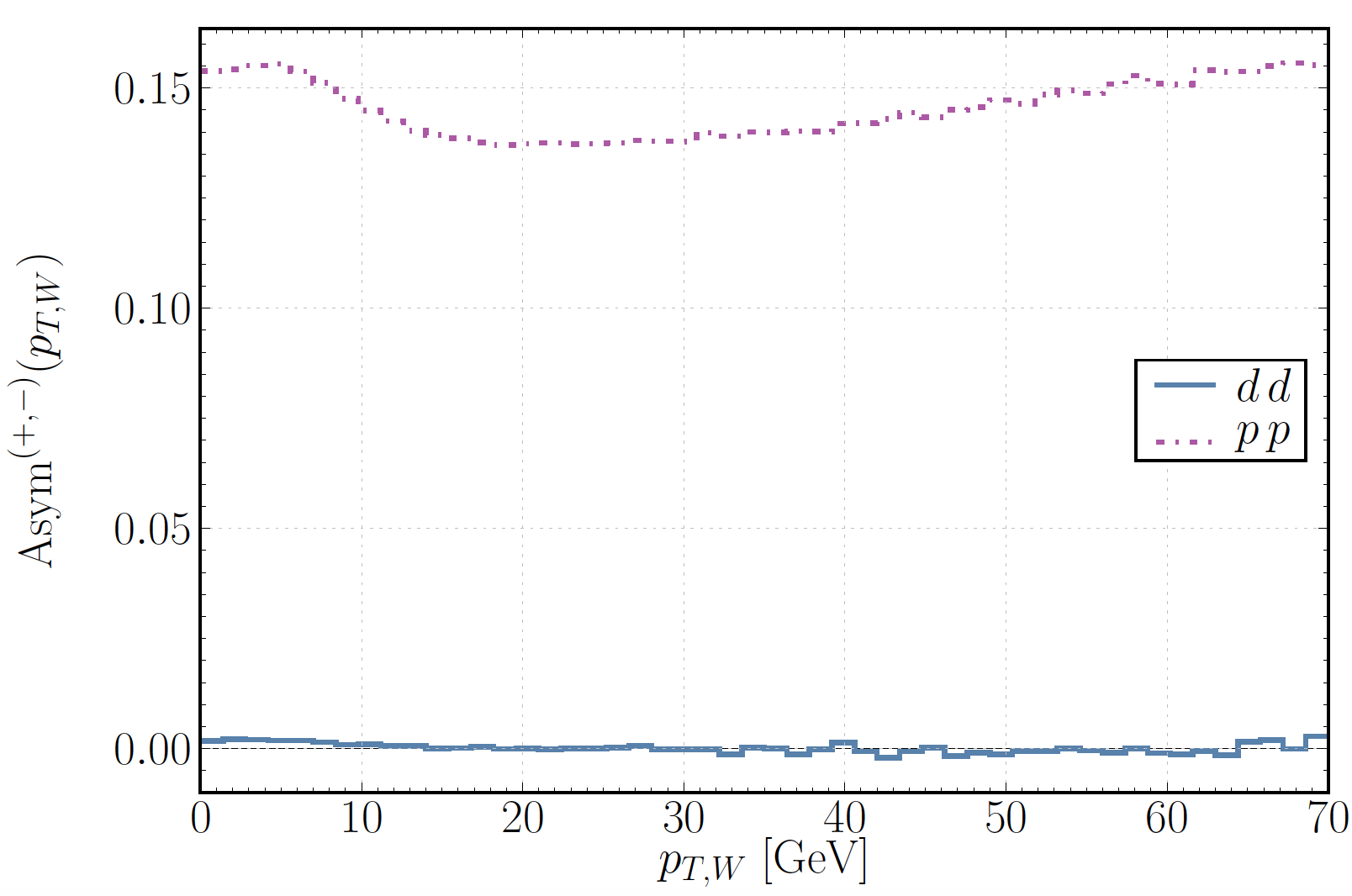}
	\includegraphics[width=0.49\linewidth]{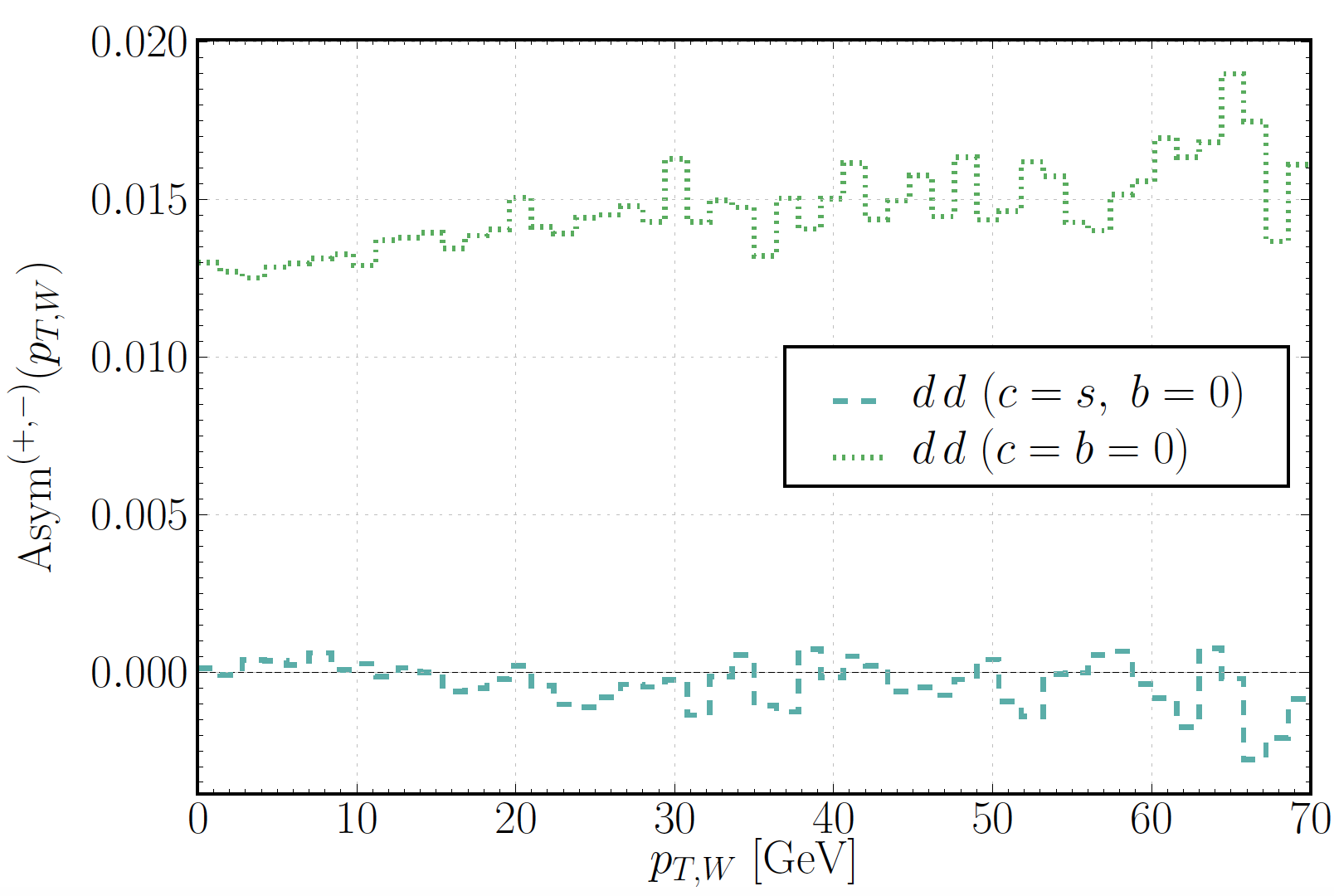}
	\caption{The predicted charge asymmetries of the transverse 
	          momentum of $W$-boson, $\pTW$,
              for the isoscalar-nuclei ($dd$) and proton--proton ($pp$) collisions
              (LHS),
              and for the $dd$ collisions under the following two assumptions: (1) $c=s$ and $b=0$, and (2) $c=b=0$ (RHS). }
	\label{W-asym}
\end{figure}

For the $pp$ collision mode, as shown in the left-hand-side plot of Fig.~\ref{W-asym}, 
the expected asymmetry is significant. Its magnitude and 
shape are predominantly driven by the asymmetry 
in the transverse momentum, $k_t$, distributions of the 
$u$ and $d$ quarks.  For the isoscalar beams, 
this dominant asymmetry source is suppressed 
and the remaining asymmetry is significantly reduced.
It is driven essentially by the Cabibbo-suppressed 
difference of the respective distributions of 
the $s$ and $c$ quarks. The contribution of the $b$ quarks 
is suppressed  by the $\Vckmsqr{u}{b}$ element of the CKM matrix.

The right-hand-side plot in Fig.~\ref{W-asym} 
illustrates the sensitivity of the $W$-boson charge 
asymmetry to the corresponding asymmetries in 
the transverse momentum distributions of 
the strange and charm quarks. Two extreme cases 
are shown, corresponding to  $c=0$ and $c=s$, 
as an illustration of  the $c-s$ constraining 
power of the  $W$-boson charge asymmetry measured 
using the isoscalar nuclear beams.

It remains to be added that in order to take the full profit from colliding the isoscalar bunches 
at the LHC, the statistical precision of the corresponding measurements must remain higher than 
the systematic one. 
This calls for the integrated nucleon--nucleon equivalent luminosity collected in such a running mode
to exceed $100\,$fb$^{-1}$ \cite{Krasny:2010vd,Fayette:2008wt}.

\subsection{New observables and measurement methods}

Isoscalar nuclear beams allow to introduce new  observables and new 
measurement methods which can be optimised to drastically reduce 
systematic errors and interpretation ambiguities 
of the LHC measurements. 
Such methods were at the heart of proposing in the 1990s,  
initially for the HERA physics programme
and later adapted for the Tevatron
physics programme, the concept of generic, 
model-independent analysis of the data collected 
at the high-energy colliders \cite{Generic}.
As an example of extension of such methods to the LHC
environment, we present below the analysis strategy of
asymmetries in inclusive charged-lepton distributions in the $Z$- and  $W$-boson production processes. 

Several subtle weak-interactions effects contribute to the 
asymmetries of the leptonic momenta distributions 
in the $Z$- and $W$-boson production processes.
The charged-current (CC) coupling of quarks to the $W$-bosons 
is of the $V-A$ type, while their  coupling to the $Z$-bosons 
is a coherent mixture of $V-A$ and $V+A$ couplings. This 
difference is reflected in the asymmetries in the angular 
distributions of leptons originating from the decays 
of the $W$ and $Z$ bosons. 
Radiative corrections affect differently the $W$ and 
$Z$-boson production and decay amplitudes.
While the effects of the QCD radiative corrections are  
driven mainly  by the  mass difference of the $W$ and $Z$-bosons, 
the effects of the EW radiative corrections lead to several 
more subtle effects. First of all, the virtual EW corrections 
affect differently the $W$ and $Z$-boson absolute
production rates. In addition, the radiation of photons 
affects differently  the $W$ and $Z$-boson 
propagation and decay. This is mainly  driven by the  
differences in the interference pattern: (a) of the amplitudes 
for the photon emission from 
each of the charged leptons in $Z$-boson decays; (b) of the amplitudes of
photon emission from the charged lepton and the charged $W$-boson.     

The analysis strategy  which amplifies the sensitivity to the above 
weak-interactions effects and drastically reduces the modelling uncertainties
driven by the partonic distribution and strong-interactions effects was proposed in
\cite{Krasny:2007cy} and is recalled below.
 
\begin{figure}[!htpb]\centering
(a)\hspace{7.5cm} (b)
	\includegraphics[width=1.0\linewidth]{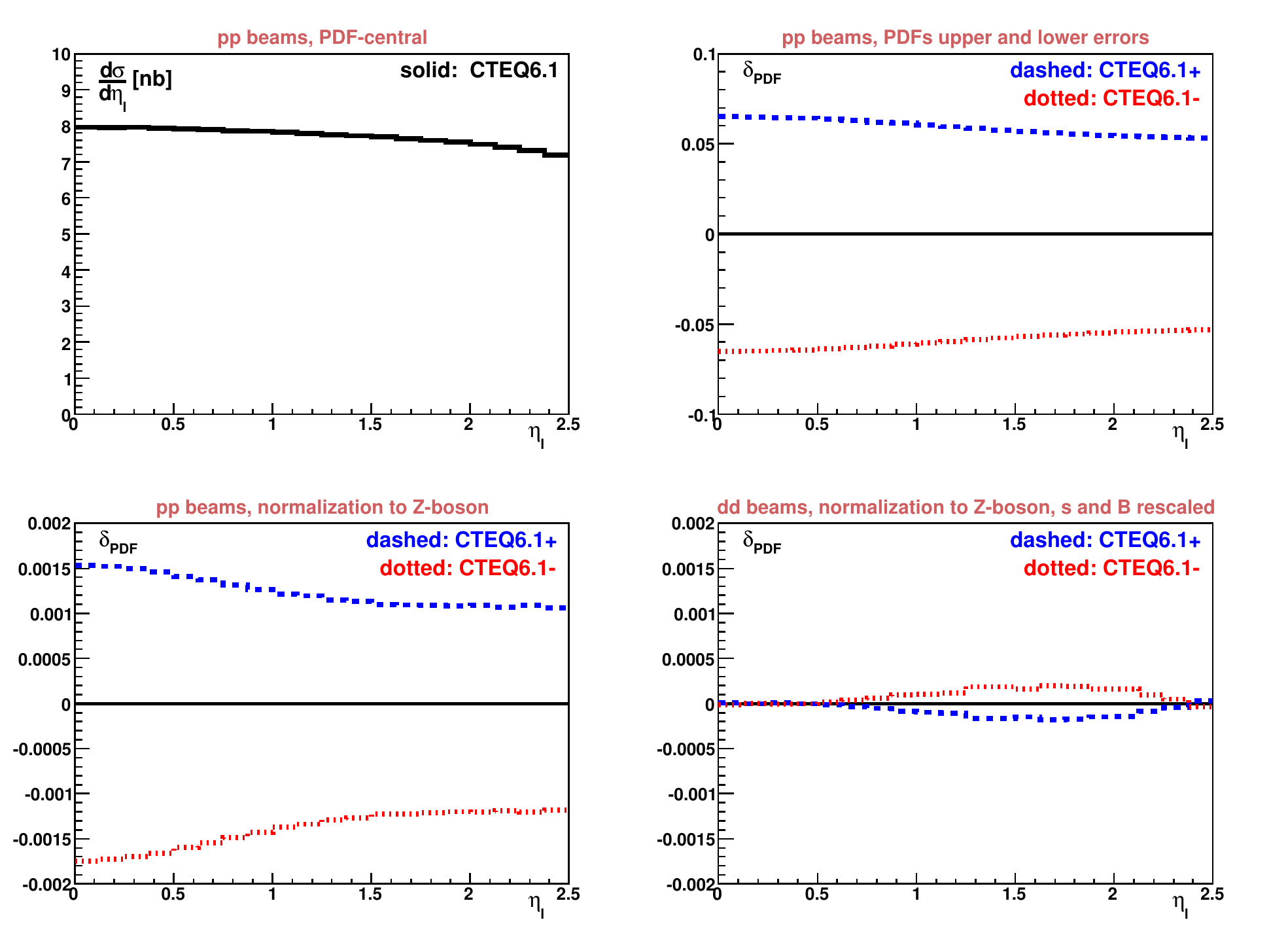}
	\caption{\sf The distribution of the lepton pseudorapidity $\eta _l$ 
for proton--proton collisions at LHC (a);  
the systematic uncertainty  
$\delta _{\rm PDF} = \frac{d \sigma/d \eta_{l}({\rm CTEQ6.1} \pm) 
- d \sigma/d \eta_{l}({\rm CTEQ6.1})}
{d \sigma/d \eta_{l}({\rm CTEQ6.1})}$ 
of the $\eta _l$ distribution reflecting  the PDF uncertainty (b);
as above but for  the ratio of the $\eta _l$ distributions for
the $W$ and $Z$-boson samples (c);  
as above but for the collision of the isoscalar beams,
re-scaled collision energy,  and re-scaled magnetic fields (see the text for 
details) (d).  }
    \label{Z-candle}
\vspace{-9.2cm}
(c)\hspace{7.5cm} (d)
\vspace{9.2cm}
\end{figure}

In Fig.~\ref{Z-candle}a  we show the charged lepton 
pseudorapidity distribution for the 
$pp \rightarrow W + X$,
$W \rightarrow l \nu$, process  
at the centre-of-mass (CM) energy of $\sqrt{s_n}= 14\,$TeV for the CTEQ6.1 parton distribution functions \cite{CTEQ6.1:2003}.
The dominant contribution to the uncertainty of this
distribution -- coming from the uncertainties in the partonic
distributions and determined using the method proposed in \cite{CTEQ6:2002} -- is shown in 
Fig.~\ref{Z-candle}b. 
The partonic distributions related uncertainty 
can be diminished to the per-mil level
by using, as an observable, the ratio of
the charged lepton pseudorapidity distributions  
for the $W$ and $Z$-boson production events.
This  uncertainty is shown in Fig.~\ref{Z-candle}c.

Further reduction of the impact of the uncertainty of the partonic
distribution functions can be achieved  
by replacing the proton collisions by the isoscalar-nucleus collisions
and by measuring a 
new observable constructed using the measurements made at the 
following two colliding beam energy settings: 
$\sqrt{s_1}$ and $\sqrt{s_2}= (M_Z/M_W) \times \sqrt{s_1}$.
These two settings  allow to keep the momentum 
fractions of the partons producing 
$Z$ and $W$-bosons equal if the $W$-boson sample 
is collected at the CM-energy $\sqrt{s_1}$ and the $Z$-boson sample 
at the CM-energy $\sqrt{s_2}$. 

This new observable is defined as
\begin{equation}
R^{\rm iso}_{WZ} = 
\frac{d \sigma^{\rm iso}_W (s_1)}{ d \sigma^{\rm iso}_Z(s_2)}\,. 
\label{eq:RWZiso}
\end{equation}
It fully preserves the sensitivity to the EW effects.
This observable is plotted in Fig.~\ref{Z-candle}d as a function 
of the lepton pseudorapidity.
Its sensitivity to the uncertainty in the partonic
distribution functions is reduced\footnote{Note that 
the nuclear effects which may affect differently 
$u$ and $d$ and quarks and the QCD-scale effects 
do not affect the observable defined above.}
by more than two orders of magnitude, from the level of 
$5 \times 10^{-2}$ to the level below  $2 \times 10^{-4}$. 

This example and the one discussed in the
previous section show that the merits of the isoscalar 
partonic bunches may be exploited in two 
complementary ways:

\begin{itemize} 
\item 
by fully constraining the flavour-dependent partonic distributions, as discussed in Section~\ref{Constraints}, 
\item 
by making these partonic distributions irrelevant for specially designed observables which amplify the EW  effects w.r.t.\ the QCD ones, as discussed above.
\end{itemize}

 \subsection{Measurement of Standard Model parameters}
 \label{SM-parameters}

The new observables and measurement methods, specially designed
for the collisions of isoscalar partonic bunches, open the path to 
high-precision measurements of the SM EW parameters,
such as the masses of the $Z$ and $W$ bosons, $M_Z$ and $M_W$, 
and the weak-mixing angle (also called the Weinberg angle) $\theta_W$. 
In SM they are inter-related  by the tree-level expression%
\footnote{This relation holds to the infinite perturbative order in the on-shell renormalisation scheme.}
\begin{equation}
\sin^2 \theta_W = 1 - \frac{M_W^2}{M_Z^2}\,.
\label{eq:sin2W}
\end{equation}
Out of the above three parameters, $M_Z$ is known experimentally to a 
much better precision than the other two \cite{PDG:2019}. 

In practice, at both electron--positron and hadron--hadron colliders,
instead of $\sin^2\theta_W$, 
the sine-squared of the so-called effective fermion mixing angle 
$\sin^2 \theta_{\rm eff}^f$ is measured. It is related to the ratio 
of the vector to axial couplings of a given fermion $f$ to the
$Z$-boson and can be expressed as (see e.g.~\cite{ALEPH:2005ab}):
\begin{equation}
\sin^2\theta_{\rm eff}^f = \kappa_f \sin^2 \theta_W,
\label{eq:sin2Weff}
\end{equation}
where the factors $\kappa_f$ account for quantum-loop corrections, in particular
those corresponding to the virtual top-quark and Higgs-boson contributions. 
Prior to the Higgs-boson discovery, the above relations were used for 
indirect determination of the SM Higgs mass. Now, when the Higgs-boson mass was  measured, SM became  over-constrained,  
and the relations (\ref{eq:sin2W}) and (\ref{eq:sin2Weff}) can be used for 
a consistency check of SM and a stringent test of various BSM scenarios. 
For example, the $\sin^2 \theta_{\rm eff}^f$ measurement can be used  as an
indirect determination of the $W$-boson mass $M_W$, which should be consistent 
with its directly measured value.

The LHC precision goal for the direct $M_W$ measurement is $\sim 5\,$MeV. 
In order to achieve a comparable precision for an indirect determination, 
$\sin^2 \theta_{\rm eff}^f$ should be measured with the precision of $\sim 10^{-4}$. 
The most precise up-to-date measurement was done by the LEP/SLD experiments
for the leptonic $\sin^2\theta_{\rm eff}^l$, with the total error 
of $1.6\times 10^{-4}$ \cite{Schael:2013ita}, which is equivalent to the indirect $M_W$-determination error of $8\,$MeV.
The  $\sin^2\theta_{\rm eff}^l$ value   has also been measured at the LHC by the ATLAS
\cite{ATLAS:2018gqq}, CMS \cite{Sirunyan:2018swq} and LHCb \cite{Aaij:2015lka}
experiments with the precision of $3.6\times 10^{-4}$, $5.3\times 10^{-4}$ and $10.6\times 10^{-4}$, respectively.
The ultimate goal of these experiments is to improve  the precision of
the LEP/SLD measurements.

The perspectives set up for the HL-LHC $pp$ operation phase by the three 
experiments \cite{ATL-PHYS-PUB-2018-037,CMS-PAS-FTR-17-001,LHCb-PUB-2018-013} 
are rather pessimistic: the final errors of the $\sin^2 \theta_{\rm eff}^l$ measurements 
will be dominated by the uncertainties 
of the momentum distributions of partons -- the corresponding errors 
are expected to be, at least, by a factor of $2$--$3$ larger than the statistical errors. 

The observable that is commonly used in experiments with
unpolarised beams, both at the lepton and hadron colliders, 
to extract $\sin^2 \theta_{\rm eff}^l$ is the forward--backward
asymmetry in the process of the charged lepton-pair production 
($e^+e^-$ and/or $\mu^+\mu^-$) near the $Z$-boson peak:
\begin{equation}
A_{\rm FB} = \frac{\sigma_{\rm F} - \sigma_{\rm B}}
{\sigma_{\rm F} + \sigma_{\rm B}}\,,
\label{eq:AFB}
\end{equation}
where $\sigma_{\rm F}$ and $\sigma_{\rm B}$ are the cross sections
in the forward and backward hemispheres. The forward (backward) hemisphere
is defined by the direction of motion of the incoming point-like 
fermions (antifermions): electrons (positrons) at $e^+e^-$ colliders 
and quarks (antiquarks) at hadron colliders.

Contrary to the $e^+e^-$ collisions where the incoming electron
direction is known, for collisions of partonic bunches
the incoming quark direction cannot be directly determined  on 
an event-by-event basis. The LHC experiments rely on statistical
correlations of this direction with the direction of the Lorentz
boost of the outgoing lepton-pair\footnote{The valence 
quarks carry on average a higher momentum fraction than sea antiquarks.}, 
which is used to  define the forward--backward hemispheres on 
the event-by-event basis and, in turn, 
the asymmetry $A_{\rm FB}$~\cite{Sirunyan:2018swq}.
 
At the LHC energy, contrary to the Tevatron case,  this observable suffers from
the  antiquark dilution corrections \cite{Sirunyan:2018swq}, requiring  a very 
precise  knowledge  of the ratio of the quark and antiquark momentum distributions.
Moreover, since the $u$ and $d$ quarks have different couplings to the $Z$-boson,
they contribute with different weights to $A_{\rm FB}$.
As a consequence,  
in the case of  the proton partonic bunches for which 
$u_v\neq d_v$, 
the $A_{\rm FB}$ observable is highly sensitive  
to the difference in the momentum distributions of the valence $u$ and $d$ quarks\footnote{One of us (MWK) is indebted to Arie Bodek 
for the in-depth discussions of the precision brick-walls 
in measuring $\sin^2\theta_W$ at high-energy hadronic colliders
and on the possible ways to overcome them.}.

For isoscalar partonic bunches 
the fractions of the valence $u$ and $d$ quarks are equal, cf.\ Eq.~(\ref{iso-constrains}),
and, as a consequence, 
the $A_{\rm FB}$ measurement no longer suffers from 
the limited precision of the $u_v - d_v$ distribution,
which drives  the measurement uncertainty for the proton bunches. 
In addition, the ratio of the quark to antiquark momentum 
distributions can be fully constrained by using the the observables 
and methods described in the previous subsection.

With the integrated luminosity of $3000\,$fb$^{-1}$ from the HL-LHC
$pp$-collision phase, the CMS and ATLAS experiments expect to 
reach the statistical errors
of the $\sin^2 \theta_{\rm eff}^l$ measurement at the level
of $3$--$4\times 10^{-5}$ 
\cite{ATL-PHYS-PUB-2018-037,CMS-PAS-FTR-17-001}, 
which is, at least,  by a factor of $\sim 4$
smaller than the uncertainty related to the limited knowledge 
of partonic momentum distributions of proton bunches. 
For the collisions of isoscalar nucleus bunches, the latter uncertainty 
is reduced to a level at which the statistical errors become dominant. 
For the integrated nucleon--nucleon luminosity of $\sim 800\,$fb$^{-1}$
collected in the isoscalar nucleus collision mode one 
would be able to reach the overall precision better than $\sim 10^{-4}$ for the $\sin^2\theta_{\rm eff}^l$ measurement per experiment.

For the direct $W$-boson mass measurement, 
the LHC experiments set an ambitious goal of reaching 
the precision of $5\,$MeV, or better.
Up to date, only the ATLAS experiment has published its first result,
with the total $M_W$ error of $19\,$MeV, 
from the 2011 LHC run at the centre-of-mass energy of $7\,$TeV
for the integrated luminosity of $4.6\,$fb$^{-1}$ \cite{Aaboud:2017svj}.
Two observables were used in this measurement: 
(1) the outgoing charged-lepton transverse momentum $p_T^l$, 
and (2) the $W$-boson transverse mass $m_T$. 
The $p_T^l$-based method is, in our view, the only method allowing 
to reach  the precision $\delta M_W < 10\,$MeV at the LHC%
\footnote{For the $m_T$-based method,  the uncertainty due 
to the missing recoil cannot, in our view,  be reduced to 
a requisite level.}. For this method,
a considerable contribution to the total error of 
the ATLAS measurement comes from
modelling of partonic distributions and QCD effects. 
The authors of \cite{Aaboud:2017svj} estimate this error 
to be $11.6\,$MeV, out of which about $70\%$
results from the partonic distribution uncertainties. 
These estimates are, 
in our opinion, rather optimistic\footnote{ 
For the detailed discussion of this aspect 
see e.g.~\cite{Krasny:2010vd,Fayette:2008wt}.}.

The studies presented in Refs.~\cite{Krasny:2010vd,Fayette:2008wt}
show that by replacing the proton beams by the isoscalar-nuclei beams 
one could  drastically reduce the above modelling uncertainties.  
Even for rather conservative assumptions on the partonic distributions 
uncertainties in the relevant $x$ and $Q^2$ regions:
$5\%$ for the valence-quarks, $10\%$ for the $c$ and $s$ quarks 
and $40\%$ for the $b$ quark, the corresponding uncertainties 
on the $M_W$ measurement can be reduced to below $5\,$MeV,
provided that a special observables and measurement procedures 
are employed. 
To reach this  precision level, the required collected nucleon--nucleon 
luminosity should be larger than $100\,$fb$^{-1}$.

Summarising, the use of the isoscalar nuclei beams 
could allow to achieve the accuracy in both the direct and indirect 
(through $\sin^2\theta_{\rm eff}^l$) $W$-boson mass measurements
at the level better than  $5\,$MeV, providing an important consistency 
test of the Standard Model as well as a stringent constraint for its possible  BSM extensions.

\subsection{New research opportunities with nuclear beams}

The LHC research programme can be 
enriched  by asking new questions and providing a 
suitable experimental framework to answer them. 
For example: 
\begin{enumerate}
\item
What is the mechanism which transmutes three 
degrees of freedom of the scalar field into the longitudinal-polarisation 
degrees of freedom of the $W$ and $Z$ bosons
in the EW-vacuum ground state? 
\item
Is there any trace of such a mechanism which could be observed 
experimentally, e.g.\ by analysing the polarisation asymmetries in
propagation of the transversely and longitudinally polarised EW bosons 
--  both in the vacuum and in the matter? 
\end{enumerate}

To address such questions experimentally, the most desirable 
tool for generic exploration of  the EW-boson sector 
would be a high-brightness polarisation-tagged beam of the EW bosons. 
If proton beams of energies exceeding $10^{17}\,$GeV were available, such beams of the EW bosons could be easily formed and used in macroscopic 
experiments --- in very close analogy to fixed-target muon-beam experiments which routinely use beams of unstable particles.
Even if experiments using beams of the EW bosons
cannot be realised at the macroscopic-length scales,
the LHC offers a reduced-scope, yet unique,  
opportunity to realise them  at the ``femtoscopic''-length scales. 

The LHC is a very efficient factory  
of the $Z/W$ bosons, producing hundreds of 
millions of them over each year of its operation. 
Their polarisation can be controlled by measuring the recoil
jet produced in their creation processes. 
The LHC unique merit is that these very short-lived
particles can be observed -- at the LHC  beam 
energies -- over a sufficiently long time  
to perform  experimental ``femtoscopic'' studies of
their properties and interactions.

The $Z/W$ weak bosons travel -- for the observers
co-moving  with the LHC bunches -- over the atomic 
distances of up to $10^{4}\,$fm  before decaying. 
This defines the maximal thickness, thus the type,  
of possible  targets which, if arranged to co-move 
with the LHC bunches,  could be employed in 
experimental studies of the properties and collisions 
of the EW bosons. Nature provides only 
one type of a target satisfying the above criteria 
-- the bunches of partons 
confined in nuclear envelopes. Their flavour 
composition and the effective target thickness 
can be tuned using the nuclear partonic bunches of the variable  
nuclear atomic mass $A$ and charge $Z$ numbers,
providing the  analysing medium of the 
$W$ and $Z$ boson properties.
The formalism and the framework of such an 
analysis is discussed in detail 
in \cite{Krasny:2005cb}. 

It remains to be added that the use of the 
nuclei as the femtoscopic-length targets for 
experimental studies of the QCD colour-confinement 
mechanism was the principal initial goal of the development 
of the electron--ion collider (EIC) concept, first for 
HERA \cite{DESY_workshop,HERA-eA,Arneodo:1996qa,Krasny:1997mv,HERA-eA-GSI}
and subsequently for RHIC \cite{Krasny:2001nin,Krasny:1999av,Snowmass_physics,Snowmass_detector,Krasny:2001tr,eRHIC,Accardi:2012qut}.
The most important condition to extend such 
studies to propagation of the EW bosons in the vacuum and the matter
is to achieve a comparable partonic luminosity in collisions
of the nuclear and proton beams.

\subsection{LHC as photon--photon collider}
\label{Photons}

The research domain where the superiority of 
the nuclear beams over the proton beams is the most evident 
is the photon--photon collision physics. 

High-energy  photon--photon collisions allow to test the Standard Model 
and to look for the presence of the BSM effects in 
a particularly clean way. In contrast to the quark and 
gluon collisions, photon--photon collisions are 
unaffected by the parasitic strong interaction 
effects. In addition, the matrix elements for photon--photon 
collisions can be predicted to the accuracy which is 
unreachable for processes involving quarks 
and gluons as initial partons.
This  allows to use the photon--photon collisions
for the absolute  measurement of the 
hadron-collider luminosity with the precision approaching 
the one achieved at the electron--positron colliders 
\cite{Krasny:2006xg,Krasny:2010vj,Krasny:2011hr}. 
Moreover, the $Z^2$-enhancement of the photon content of the nuclear 
bunches, discussed in Section~\ref{photons}, is driven 
by peripheral processes in which the recoil nucleus does contribute to the energy deposition in the LHC detectors  --- the photon--photon collisions
are thus not obscured by the collisions of spectator partons.

\begin{figure}[!htpb]\centering
	\includegraphics[width=0.75\linewidth]{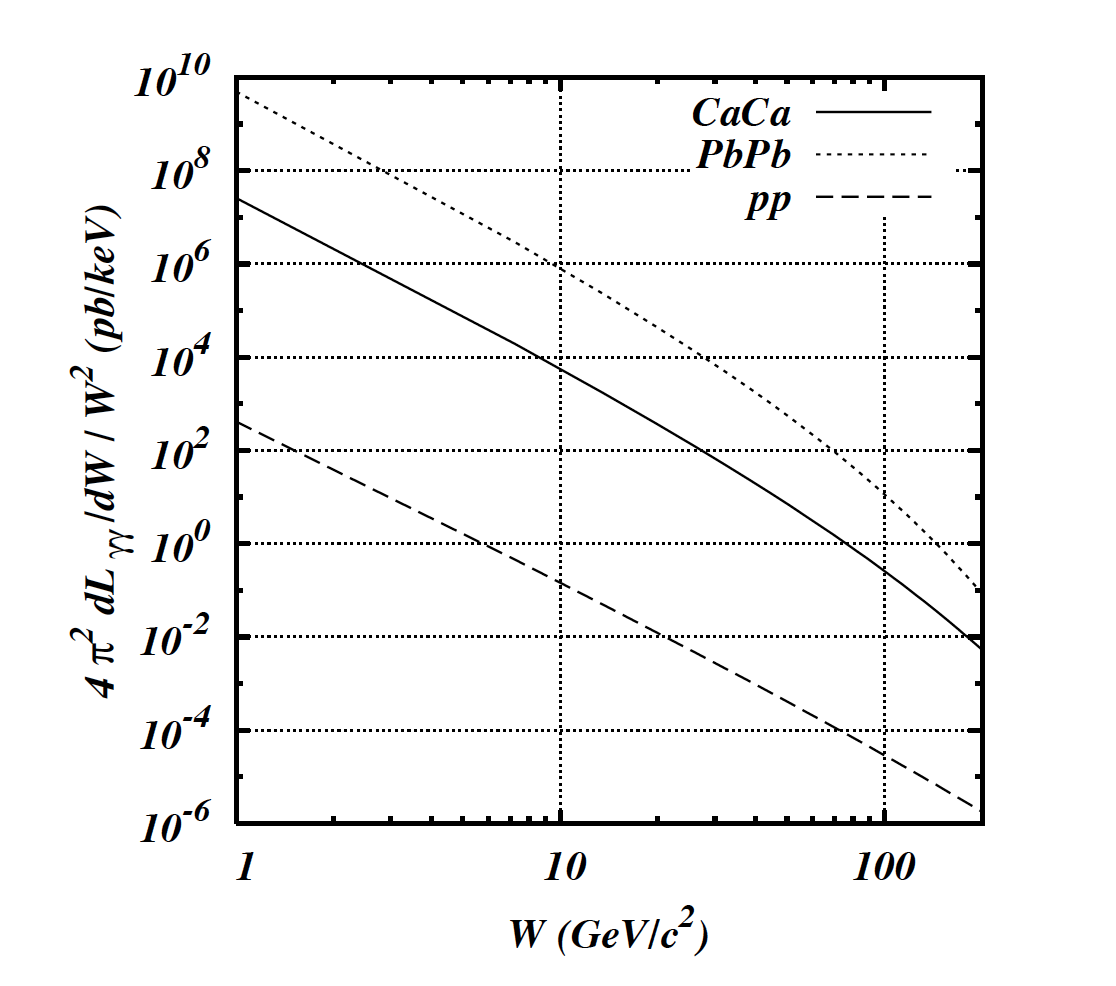}
	\caption{The effective luminosity for the resonance production in the photon--photon collisions as a function of the photon--photon collision centre-of-mass energy for the $pp$, $\mathrm{CaCa}$ and $\mathrm{PbPb}$ collision modes (for more details see \cite{Baur:2001jj}).}
    \label{gamma-gamma-resonance}
\end{figure}

To illustrate the advantage of the nuclear bunches for 
studies of photon--photon collisions, in
Fig.~\ref{gamma-gamma-resonance} we show the effective luminosity 
$4\pi^2dL{\gamma \gamma}/dW/W^2$ for the resonant photon--photon
collisions at the LHC with the CM energy, $W$
for the $pp$, $\mathrm{CaCa}$ and $\mathrm{PbPb}$ collisions \cite{Baur:2001jj}. 

The effective luminosity enhancement factors for the nuclear beams 
are very large --  even if scaled down to the same nucleon--nucleon
luminosity (the $A^2$ scaling), they are at the level of $200$ for the $\mathrm{PbPb}$
collisions and of $60$ for the $\mathrm{CaCa}$ collisions. 
They open several new research options at the LHC, including 
searches of axion-like dark-matter particles (ALPS), 
precision studies of the elastic light-by-light scattering 
and Higgs-boson production processes. 
The latter of these options is discussed in the next section.

\subsection{Exclusive Higgs-boson production }
\label{Higgs}

Perhaps the most remarkable  physics highlight of
high-luminosity collisions of nuclear beams at the LHC 
would be to observe the exclusive production of the Higgs bosons 
in the photon--photon collisions and their background-free decays.
Let us choose,  for the discussion presented below, the case 
of the isoscalar calcium ($\mathrm{Ca}$) beam. 

In the high-luminosity proton--proton collision phase,  
the integrated luminosity of the order of $3000\,$fb$^{-1}$ is expected to be delivered to the LHC experiments.  
If the same  nucleon--nucleon luminosity can be achieved 
for collisions of the calcium beams and $30\%$ 
of the running period is devoted to
such a collision mode, the delivered  $\mathrm{CaCa}$ 
luminosity would
be $625\,$pb$^{-1}$. The expected  number of exclusively
produced Higgs bosons, in such a scenario,  would 
be $N_{\rm Higgs} \approx  420$ per
experiment. In total $240$, $90$, $26$ and $10$ Higgs-boson
decays to the $b\bar{b}$, $WW^*$, $\tau \tau$ and
$ZZ^*$ pairs, respectively, could be observed by each of the LHC experiments.

The above  numbers are smaller, by large factors,  than the 
corresponding numbers for the  gluon--gluon collisions. 
However, the $\mathrm{CaCa}$
running mode could provide the first experimental evidence of the $s$-channel, exclusive, resonant 
Higgs-boson production in photon--photon collisions. Such 
an evidence  would strengthen the interpretation of 
the $125\,$GeV-mass peaks observed in the gluon--gluon scattering
channel as originating from the SM Higgs-boson
decays. More importantly, the decays of the exclusively produced 
Higgs bosons can be easily identified and measured
with negligible background\footnote{It is important to note
here that the expected number of  $\mathrm{CaCa}$ collisions per bunch crossing is reduced with respect to the high-luminosity proton-beam collisions by a very large factor,  allowing to observe exclusive 
photon--photon collision events without the need for very 
forward ion taggers. This aspect is discussed in more details
in the subsequent section.}.
This could facilitate the detection 
and the measurement of the $H \rightarrow b\bar{b}$ decay mode --  
the  mode which is difficult to detect and measure in the gluonic
collisions of the beam particles because of too a large 
irreducible background. According to our estimates, 
if only one twentieth of the $pp$ running time in the
high-luminosity mode would be  attributed 
to the  $\mathrm{CaCa}$ collisions, the  
$\gamma \gamma \rightarrow H \rightarrow b\bar{b}$ process could 
be discovered with more than the 5$\sigma$ evidence. 
For more complete studies  of the $\gamma \gamma \rightarrow H \rightarrow b\bar{b}$ discovery potential using nuclear beams at the LHC see e.g. \cite{dEnterria:2019jty}.

Finally, let us add that the Higgs production cross-section
and photon fluxes increase very fast with increasing energy 
of the ion beam. For the high-energy LHC, with the doubled beam energy, 
the expected number of produced Higgs particles increases  
by a factor of $\approx$ 10, opening the possibility 
of a competitive precision measurements of the Higgs-boson couplings.

\subsection{QCD studies} 

The discussion of the relative merits of the 
proton and nuclear beams, presented so far, has been restricted to
the SM EW and BSM sectors. For these sectors, the
measurement of the beam particles' partonic distributions
can be considered as the development of the specialised analysis 
tools, or even as an analysis burden, rather than a subject of
dedicated studies. For the QCD sector, on the contrary, the 
measurement of partonic distributions is of pivotal importance. 

Partonic distributions provide a unique insight into the mechanism of  
strong interactions of quarks and gluons which confines them  in the 
nucleons and nuclei. They can be used for quantitative  tests of  the 
QCD predictions in the  kinematical domains where 
the perturbative computational  methods can be applied. 
More importantly, they are crucial for phenomenological studies 
of the strong interactions at large distance scales  
which cannot be, at  present,  controlled by the available 
QCD computational methods. 
The merits of nuclear partonic bunches are of particular importance 
for these studies. 

The data collected so far at the LHC in the low-luminosity 
$p\mathrm{Pb}$ and $\mathrm{PbPb}$ collision runs
for the Drell--Yan production of on-shell and off-shell EW bosons --  
which could potentially probe partonic distributions with high  
precision -- are suffering from large statistical errors. The LHC nuclear 
collision data have, so far, hardly added any new knowledge 
to that acquired from the analysis of the fixed-target data collected 
in the pre-LHC era \cite{Paukkunen:2018kmm}.

The  high-luminosity collisions of nuclear bunches 
would open new perspectives for the QCD studies at the LHC. 
Their highlights include:
\begin{itemize}
\item
The validity test of the QCD DGLAP equations\cite{DGLAP} 
describing the $Q^2$-evolution of 
partonic distributions could be extended to the ``high partonic
density regime'' involving partons carrying a  small 
fraction of the nucleus momentum. 
\item 
Studies of the colour conductivity effects \cite{Reya} in nuclei, 
by comparing the measured partonic distributions in 
their central and peripheral collisions.
\item
Precise measurement  of the QCD coupling constant $\alpha_s$,
by  measuring the inclusive spectra of leptons 
produced in collisions of isoscalar nuclei \cite{Krasny:2007cy}.
\item
Studies of  the relative density  of protons and neutrons
in the centre and on  the surface of the nucleus, and the 
corresponding studies of the 
neutron skin effect \cite{Paukkunen:2015bwa},
by measuring the asymmetries in the 
transverse momentum distributions of positively and negatively 
charged $W$-bosons in collisions of isoscalar beams.  
\item
Studies of the mechanism of binding nucleons within nuclei, 
by precision measurement of the $R^A _{v,s}$ factors (see Eq.~(\ref{eq:RAq})) 
in the wide domain of $x$ and $Q^2$ variables, with the significantly 
higher precision than that  achieved so far.
\end{itemize}

Last but not least, the high-luminosity collisions of light nuclei 
could provide a new environment to study 
the quark--gluon plasma signatures, 
by enabling the observation of rare collision events with abnormally 
high particle multiplicities. Such studies would
complement the studies of the quark--gluon plasma 
signatures in the low-luminosity collisions of heavy nuclei.

 \subsection{Pile-up background}
 \label{Pileup}

The precision of the EW measurements and the sensitivity 
of searches for the BSM processes in the high-luminosity 
$pp$-collision phase of the LHC operation will suffer from  
the large  pile-up background. Already in the present phase
of the LHC operation with proton beams,  a collision event of 
interest is  accompanied on average by $\nu = 30$ background pile-up 
collision events occurring  within the time of the  bunch crossing 
of $\approx 700\,$ps. 
They give rise to energy depositions 
in the LHC detectors which, in a majority of cases, cannot be unambiguously
attributed to the signal and background events. In addition, they
produce a large number of hits in the LHC detectors trackers
-- a severe problem for  fast track-finding algorithms affecting    
the trigger and event selection  efficiency. In the high-luminosity 
LHC phase, with the $50\,$ns interval between the beam bunches,  
$ \nu$ is expected 
to rise to the value of $454$~\cite{Apollinari:2017cqg}.
Since the LHC detectors cannot cope with such a pile-up rate, 
the LHC luminosity will have to be levelled to value of 
$2.5 \times 10^{34}\,$cm$^{-2}$s$^{-1}$.
For this luminosity, the pile-up will be reduced to $\nu = 150$ 
-- the value that is sustainable for 
the LHC-detectors operation. For the alternative high-luminosity LHC operation 
with the $25\,$ns bunch-spacing, the levelled luminosity 
is expected to be twice larger (at the same $\nu$ value). 

A very important merit of nuclear beams is that the 
number of collisions per bunch crossing can be drastically  
reduced while preserving the same nucleon--nucleon luminosity 
in the $AA$ collision mode as in the  $pp$ one. 
The relationship between $\nu _{AA}$ and $\nu _{pp}$
can expressed, for the same nucleon--nucleon luminosity,  as follows:
\begin{eqnarray}
 \nu _{AA} = \nu _{pp} \times \frac{\sigma _{AA}}{\sigma _{pp}} \times A^{-2},
\end{eqnarray}
 where $\sigma _{AA}$ and $\sigma _{pp}$ are the respective 
 inelastic cross sections of the $AA$ and $pp$ collisions, respectively.
 The number of collisions per bunch-crossing decreases with the increasing atomic number proportionally to $A^{-4/3}$ --  by a factor of $40$, $136$, $650$ and $1260$
 for the $\mathrm{OO}$, $\mathrm{CaCa}$, $\mathrm{XeXe}$ and $\mathrm{PbPb}$ collisions, respectively. Such a large reduction factors 
 open the possibility of observing exclusive peripheral collisions of nuclei 
 at the high nucleon--nucleon luminosity in a pile-up free mode.

 The decrease of the $\nu$ value is, of course, associated with an  increase of  multiplicity of particles produced in each of the $AA$ collisions. 
 However, at the {\it high nucleon--nucleon luminosity}, the average total
 multiplicity of background particles per bunch-crossing produced in the fiducial
 volume of the LHC trackers turns out to be  smaller for the $AA$ collisions than 
 for the $pp$ collisions (for the same effective nucleon--nucleon luminosity). 

The above  statement is, at first sight,  counter-intuitive
and requires further explanation. 
For the low nucleon--nucleon luminosity, defined by the condition 
$\nu_{pp} \leq 1$, the opposite is true because the number of
background particles produced in a single $AA$ collision is significantly higher than
that in a single $pp$ collision. 

\begin{figure}[!htpb]\centering
	\includegraphics[width=0.75\linewidth]{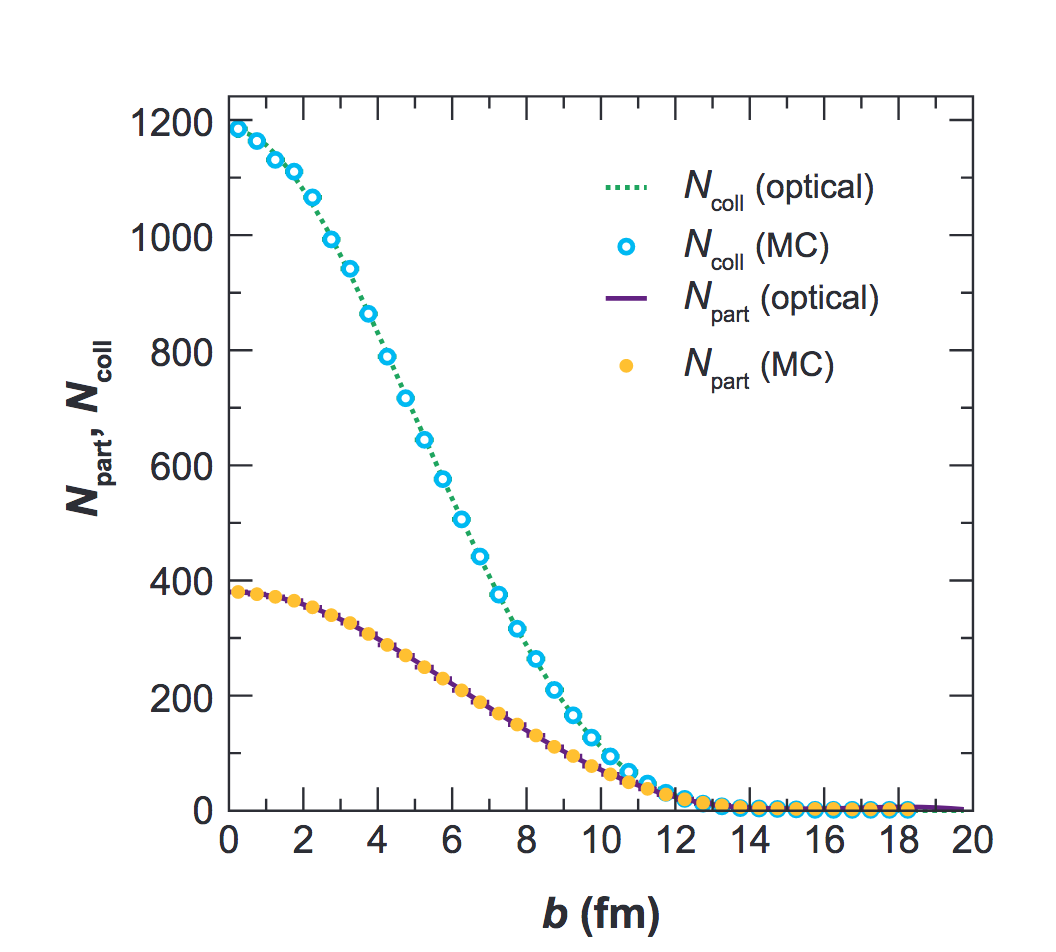}
	\caption{The number of binary collisions $N_{\rm coll}$ and the number 
	          of collisions participants $N_{\rm part}$  as a function 
	          of the impact parameter, calculated in the optical
	          approximation (lines) and with a Glauber Monte Carlo
              (symbols) for the $\mathrm{AuAu}$ collisions, taken from
              \cite{Miller:2007ri}. Note that $N_{\rm coll}$
              is a measure of the nucleon--nucleon luminosity, both for 
               the $pp$ and $AA$ collisions. The ratio $N_{\rm part}/N_{\rm coll}$
               is thus the expected reduction factor of particle
               multiplicity in the central region in the $AA$ collisions
               with respect to the $pp$ collisions. }
    \label{N_part}
\end{figure}

At high luminosity, when $\nu _{pp}$ is larger than the atomic mass number  $A$ of
the colliding nuclei, it  matter less  whether 
colliding partons are delivered to the interaction point 
in the proton or in the nuclear envelopes. The
multiplicity of background particles per bunch-crossing is expected 
to be driven by the number binary collisions, $N_{\rm coll}$, 
of free (the proton case) and bound (the nuclear case) nucleons. 
The bound nucleons undergo soft collisions 
several times over their passage through the volume of 
the target nucleus. They ``loose'' their soft partons 
already in their first collision and produce 
less particles per binary collisions than free nucleons 
(protons)\footnote{One of us (MWK) is indebted to Wit Busza 
for a discussion of the multiplicity of particles produced 
in $pp$ and $AA$ collisions.}. For nuclear bunches, 
the number of produced particles is no longer
proportional to the number of binary collisions but, instead, 
to the number of pairs of wounded (participant)
nucleons \cite{Bialas:2012dc}, $N_{\rm part}/2$. The difference of 
$N_{\rm coll}$ and $N_{\rm part}$ for the $\mathrm{AuAu}$ collisions
is  shown if Fig.~\ref{N_part}. $N_{\rm part}/2$, driving 
the background level for $AA$ collisions, is significantly 
smaller than $N_{\rm coll}$, determining -- for the equivalent 
nucleon--nucleon luminosity -- the background level in the $pp$ 
collisions. The background reduction is expected to be particularly 
significant in the central $AA$ collisions characterised by 
a small impact parameter $b$ of the colliding nuclei. 
On average, we expect 
the reduction of the pile-up background level for the gold ($\mathrm{Au}$) beams 
by a factor of $4$ 
as compared to the one for the  $50\,$ns bunch-spacing $pp$ operation mode 
and by a factor of $2$ as compared to the $25\,$ns one. This would
allow to operate the LHC detectors at higher nucleon--nucleon 
luminosity by using the nuclear rather than proton partonic bunches.

\section{Towards high-luminosity $AA$ collision scheme}
\label{Luminosity}

\subsection{Introductory remarks}
\label{remarks}

In order to consider seriously a scenario of sharing  
the LHC running time
between  the $pp$ and $AA$ collision modes in the 
high-luminosity  phase of its  operation, and to 
fully profit from the numerous 
merits of nuclear beams, put in light in the previous 
section, a new $AA$ 
beam-collision scheme has to be proposed which satisfies 
the achievable luminosity requirement specified by the following 
condition:
\begin{eqnarray}
L_{AA} = L_{NN}/A^2 \approx L_{pp}/A^2,
\label{lumi1}
\end{eqnarray}
where:  $L_{AA}$, $L_{NN}$ and  $L_{pp}$ are, respectively,  the nucleus--nucleus,  
effective nucleon--nucleon and $pp$ luminosity in the high-luminosity 
phase of the LHC operation. This condition 
assures a comparable  partonic luminosities for the $pp$ and $AA$ 
collisions and, as a consequence, both the equivalent  statistical 
precision of the EW measurements and the equivalent sensitivity 
to the BSM effects\footnote{As discussed in Section~\ref{energy_frontier}, 
the equivalence of the $pp$ and $AA$ collision modes 
is restricted to the kinematical domains which are accessible to both 
collision modes and where the partonic-luminosity penalty factors are close to $1$.}.

The concept of such a scheme will be presented in this section. 
Prior to its discussion a more general comment is, however, mandatory. 

The $pp$ collision mode is the only one to search for the BSM effects 
at the highest partonic-collision CM energy 
$ \sqrt{s} \geq 3 $ TeV,  not accessible for the $AA$ collision mode. 
It is also clearly superior for the precision measurements 
of the EW processes for which 
the cross section is a strongly rising function of $\sqrt{s}$,
such as e.g.\ the production of triplets of the EW bosons. 
Therefore, it will always remain the principal running mode 
of the LHC. However, if the $AA$ collision luminosity 
target, specified above, can be achieved, then devoting of the order 
of 30\% of the $pp$ running time to the $AA$ mode would
have a negligible impact of the $pp$ collision results, 
while opening new research domains specific 
to the $AA$ collision mode. 

The above argument is strengthen by 
the lack of evidence of any BSM anomaly in the $pp$-mode exclusivity 
domain in the data collected so far at the LHC. That is why, 
in our view, complementary LHC running modes should be 
studied and -- if their viability is experimentally 
proven -- incorporated in the LHC operation planning.

\subsection{Optimisation parameters  }
\label{parameters}

The nucleon--nucleon luminosity in the $pp$ and $AA$ collisions
of bunched beams can be specified by the following expression:
\begin{eqnarray}
L_{NN}= f_0 \gamma_L n_b\, \frac{N_N^2}{4 \pi \epsilon _n \beta ^*} 
\,H\left(\frac{\sigma _z}{\beta ^*}, \theta _c\right),
\end{eqnarray}
where $f_0$ is the beam revolution frequency, $\gamma_L$ is the 
beam Lorentz factor, $n_b$ is the number of colliding bunches, 
$N_N$ is the number of nucleons in the bunch ($N_N = N_p$ for 
the proton bunches and $N_N = A N_A$ for the nuclear bunches), 
$\epsilon _n$ is the normalised transverse emittance of the beams, 
$\beta ^*$ is the value of the beta function at an interaction point and  
the function $H(\sigma _z/\beta ^*,\theta _c)$
describes the geometrical luminosity reduction -- driven by a  
non-zero beam crossing angle $\theta _c$ and by the hourglass 
effect. 
For the definitions of the accelerator physics terms and parameters used in this paper see the very recent review ``Modern and Future Colliders" by Vladimir Shiltsev and Frank Zimmermann \cite{Shiltsev:2019rfl}
and references quoted therein.

In the following we shall assume the same interaction-point 
(IP) optics, the small beam crossing angle and the same number 
of bunches for the proton and nuclear  beams\footnote{The   
high-luminosity LHC $pp$ operation mode with $50\,$ns bunch spacing 
foresees $1404$ bunches   \cite{Apollinari:2017cqg} 
while the $AA$ mode $1232$ bunches \cite{Citron:2018lsq}.}.
Under these assumptions the ratio of the $NN$
and $pp$ luminosities can be expressed as

\begin{eqnarray}
\frac{L_{NN}}{L_{pp}}= \frac{\gamma _L^{A}}{\gamma _L ^p} \times 
\frac{A^2 N_A^2}{N_p^2} \times \frac{\epsilon _n^p}{\epsilon_n^A}.
\label{lumi} 
\end{eqnarray}

For the HL-LHC $pp$ operation mode one expects:
$N_p = 3.5\times 10^{11}$ ($50\,$ns bunch bunch separation) 
and $\epsilon_n^p = 3\,$mm$\,$mrad \cite{Apollinari:2017cqg}.
The expected $pp$ operation 
peak luminosity (without a crab-cavity) is 
$L_{pp} = 8.4 \times 10^{34}\,$cm$^{-2}$s$^{-1}$. 
Achieving a comparable $NN$ luminosity in 
$AA$ collisions without modifying the beam collision optics
is anything but easy. 
It will require increasing of the nuclear-bunch population 
$N_A$ or reduction of the nuclear-beam transverse emittance, 
$\epsilon _n ^{ion}$, or both. 
The present  and the target values for the number of nuclei per bunch 
and the transverse emittance of the nuclear beams at the LHC are discussed 
in the following section.

\subsection{Bunch intensity}
\label{bunch-intensity}

The quest for the maximum intensity of the nuclear (ion) beams is 
limited by many inter-correlated effects 
which affect both the single-beam-particle and bunch 
dynamics.  The bunch-intensity limit in the LHC 
injectors depends strongly on: (1) the technical, $Z$-dependent  
constraints of the present LHC injectors; (2) the 
stripping stages 
of the ion beam which must be optimised to circumvent 
the bunch space-charge constraints; (3) intra-beam scattering 
constraints; (4) beam losses due to electron stripping 
in the collisions of the beam particles with 
the residual gas in the LHC injector rings 
and (5) several other less important effects. The precise assessment 
of the achievable  $N_A$ values can be done 
on ion-by-ion bases by empirical optimisation of 
the operation mode of the ion source and the full 
LHC injection chain.

The deuteron beam, an obvious  candidate for 
the research programme  advocated in Section~\ref{Merits}, 
cannot meet the luminosity requirement expressed 
in  Eq.~(\ref{lumi1}) for the reasons explained below. 

The dedicated studies presented in Ref.~\cite{Stovall} 
showed that Linac4~\cite{Arnaudon:2006jt}, providing highly efficient
acceleration of the $\mathrm{H}^-$ ion beam\footnote{Protons for the 
high-luminosity LHC operations are initially accelerated 
as $\mathrm{H}^-$ ions, carrying two attached electrons. Following the 
Linac4 acceleration phase the electrons are stripped off.}, 
is not a feasible candidate for accelerating $\mathrm{D}^-$ ion beam. 
The reason is that, while providing good longitudinal 
acceptance 
for ions injected at a half of the proton injection velocity, 
the transverse 
RF-defocusing of the beam would be too strong, resulting 
in the excessive transverse beam losses. 
This leaves Linac3~\cite{Benedetti:2018hbx}, optimised 
for  heavy ions, 
as the only option to accelerate the deuteron beam. 
The initial estimate  presented in Ref.~\cite{AbelleiraFernandez:2012ni} 
shows that the achievable $N_A$ value for deuteron bunches would 
be in the range $0.3 \times 10^{10}  \leq N_D \leq 1.5 \times 10^{10}$.
The expected number of nucleons per deuteron bunch would thus be by 
a factor of $12$--$60$ lower than that for the proton beam.
If the beam emittance stays  the same for both beams, 
the $L_{NN}$ value for deuteron--deuteron collisions at the LHC would 
be by a factor of $288$--$7200$ lower than the $L_{pp}$ value. 
Such a luminosity drop cannot, 
obviously, be compensated by the corresponding reduction of the beam
transverse emittance.  The  deuteron-beam collision option is thus  
discarded from the further discussion presented in this paper. 
A similar conclusion is valid also for
the helium beam, for which a  poor transmission in Linac3 
was reported in Ref.~\cite{Hill:2001vt}. 

The present knowledge accumulated at 
CERN while running $Z \geq 8 $ ion beams can be expressed 
by the empirical formula \cite{Citron:2018lsq} relating 
the ion-bunch intensity for an arbitrary nucleus to 
the achieved value for the $\mathrm{Pb}(A=208,Z=82)$ 
bunches, $N_{\rm Pb} = 1.9 \times 10^8$:
\begin{equation}
 N_A(Z,A) = N_{\rm Pb} \times \left(\frac{Z}{82}\right)^{-1.9}.
\end{equation}

The expected numbers of nucleons per bunch according to 
the above formula are smaller than that for the proton 
beam: by a factor $1.4$ for the oxygen, $3.2$ for the calcium, 
$6.5$ for the krypton and $8.8$ for the xenon  bunches. 
Thus, to achieve the comparable $NN$ luminosity 
in the $AA$ collision mode as that in the $pp$ mode with 
the present CERN ion-beam source and the LEIR ring, the only 
way forward is to try to compensate the bunch 
population decrease (w.r.t.\ the proton bunches) by 
the corresponding decrease of  the transverse beam emittance. 
While for the highest-$Z$ beams this is very hard, if not impossible, 
to achieve, for the low-$Z$ ones, such as  the  beams of 
oxygen and calcium, it is worth trying.

\subsection{Beam emittance}

Two questions  have to be addressed while considering 
the high-luminosity option of the LHC with the 
low-emittance nuclear beams. The first is how to reduce 
the transverse beam emittance of colliding beams. 
This will be discussed in the next section. The second, 
discussed below, is  how to preserve the beam emittance 
in the presence of the intra-beam scattering. 

Let us assume, for a while, the same IP lattice 
parameters, the same number of nucleons and the same bunch  
emittances for  the proton and ion  beams colliding at their 
maximal energies  related  to each other by 
$\gamma_L^A = (Z/A) \gamma_L^p$.
The relationship of the emittance-growth parameter,
defined as $ \alpha _{\rm IBS} = 1/\epsilon \times d\epsilon/dt$, 
for the ion and proton beams is well approximated 
by\footnote{One of us (MWK) is indebted to John Jowett for illuminating discussions 
on the emittance growth rate in the CERN accelerator rings.}

\begin{eqnarray}
 \alpha _{\rm IBS}^A = \frac{Z^3}{A^2} \times \alpha _{\rm IBS}^p.
\end{eqnarray}

The emittance growth is stronger for the ion beam than 
for the proton beam for the same number of nucleons 
in their respective bunches:  by a factor of $2$ for the oxygen, 
$5$ for the calcium, $9.5$ for the xenon and $12.7$ for the lead bunches. 
However, for the realistic $N_A(Z,A)$, discussed in 
the previous section, the emittance growth for
the proton and for the ion beams become comparable -- they 
are larger for the latter by a factor
of: $1.4$ for the oxygen, $1.6$ for the calcium, $1.5$ for the xenon and 
$1.4$ for the lead bunches.

For the normalised emittance of  the proton beam 
in the high-luminosity phase of the LHC operation of $3\,$mm$\,$mrad, 
the expected emittance growth rate is  $17.2$ hours 
\cite{Apollinari:2017cqg}. There is thus 
a room to compensate  reduction of the number 
of nucleons in the ion-beam bunches by colliding the ions beams with 
the reduced transverse beam emittance, w.r.t.\ to that for the proton beam,
within tolerable emittance growth boundaries. 
This is the path that will be explored in the following.

\subsection{The proposed scheme}
\label{the-scheme}

The emittance of the bunched beams is determined by     
the beam-source emittance and by its growth over the bunched-beam 
acceleration and the storage time. To reduce their transverse emittance, 
the beams have to be cooled. There are several methods 
to cool ion beams. In the initial stage of 
ion beam acceleration (LEIR ring)  the electron cooling 
method is used. The corresponding reduction 
of the beam emittance is, however, insufficient 
and done ``too early" in the beam acceleration cycle
to be preserved over the subsequent phases of the beam acceleration 
in the PS and SPS rings. The LEIR cooling must thus 
be  complemented by another cooling method, optimally 
in the advanced stage of the beam acceleration 
process\footnote{With the increase of the beam energy,
the rate of the intra-beam scattering (IBS) decreases proportionally 
to the Lorentz $\gamma_L$-factor, reducing significantly the 
``post-cooling-phase"
emittance growth.}.

In this paper we propose to 
reduce the beam transverse emittance by laser cooling,
exploiting {\it the atomic degrees 
of freedom} of the beam particles\footnote{Such a cooling 
method can be applied only for the ion beams -- the proton beams 
are excluded  because $H^-$ ions cannot be accelerated
in the CERN circular machines.}. 

The final stripping of the electrons 
attached to their parent nuclei, being routinely done in 
the PS--SPS beam-transfer line is, in the proposed scheme, 
postponed and follows the beam cooling phase. 
Ideally, the optimal phase for the beam cooling would 
be when the beam is accelerated to its top LHC energy. 
Unfortunately, the subsequent collisions 
of the atomic beams in  the LHC interaction points would lead to too 
a high rate of the electron stripping in beam--beam 
collisions -- the lifetime of such beams would be 
lower than a couple of seconds \cite{Krasny:2004ue}. 
Therefore, we propose to cool the atomic beam 
at the top SPS energy and, subsequently, to strip  the remaining 
electron(s) in the transfer line between the SPS and the LHC.
The fully stripped ion beam is then accelerated 
to its maximal energy at the LHC. In such a scheme,
a care should be taken to reduce, as much as possible, 
the emittance growth in the LHC bunch collection 
and acceleration phase (e.g.\ by the controlled 
longitudinal emittance blow up during the bunch 
collection and ramping time), and, if necessary, to reduce  
the emittance growth over the beam collision run by the 
conventional stochastic cooling method 
\cite{Blaskiewicz:2007zza,Blaskiewicz:2008zz,Schaumann:2014lba},
or by the optical stochastic cooling method \cite{Lebedev:2014cha}.

\section{Laser cooling of ultra-relativistic atomic beams}
\label{Cooling}

Laser cooling is a well-known technique in the atomic 
physics. It has been successfully used to cool beams of weakly relativistic 
ions \cite{Schroder:1990zz,Hangst:1991zz} and there are 
plans to use this technique by the GSI FAIR project \cite{Winters:2015bds}
for  moderately relativistic ones. 
At the SPS, where  the maximum Lorentz factor of the ion beam 
reaches the value of $\gamma_L \approx 200$,
the ions are highly relativistic. Cooling 
of highly relativistic ion beams has not been 
experimentally studied so far.

The laser cooling is based on resonant absorption of 
the laser photons by the beam particles and by subsequent 
emission of photos.
The cross section for this process is very large if the 
energy of the photons is tuned to the resonant atomic transitions of the 
beam particles. Therefore, to achieve the fast and efficient 
cooling, the ions have to carry, during the beam-cooling 
phase, a fraction of their attached electrons. 
Such ions are,  in the following, called partially stripped ions (PSIs). 
The kinematics and dynamics 
of the photon absorption and emission process is presented 
in Appendix~\ref{annex:kinematics}.

The PSI energy loss due to emission of the photon is very 
small as compared to its energy. However, 
over multiple turns in the 
storage rings, even a small energy loos 
can significantly influence the PSI-beam 
dynamics. The laser cooling mechanism is similar 
to that of the synchrotron-radiation cooling for 
the electron beam. The most notable difference
is that while the latter is a spontaneous (random)
process, the former can be stimulated and precisely 
controlled by the suitable tuning  of the laser pulse parameters, 
such as its power, the photon-wavelength spread and offset, 
the photon transverse spot size and its offset with respect to 
the ion-beam spot. This allows to selectively manipulate a chosen 
fraction of ions within their bunches with an unprecedented precision.

 \subsection{Longitudinal cooling}
 \label{Longitudinal-cooling}
 
Energy loss of the PSI, absorbing and re-emitting the photon, 
grows as $\gamma _L ^2$. 
Since the higher-energy ions loose energy faster than the 
lower-energy ones, this naturally leads to the reduction of 
the energy spread of the PSI beam \cite{BessonovKimPRL1996}.

In the bunched beam, the PSI energy 
experiences synchrotron oscillations (around the central energy) 
which are coupled with the longitudinal oscillations. Therefore, 
the reduction of the energy spread also leads to the reduction 
of the longitudinal bunch size.
The rate of such cooling is slow -- the typical cooling time is equal 
to the time it takes to radiate the full ion energy. 

\begin{figure}[!htpb]\centering
	\includegraphics[width=0.95\linewidth]{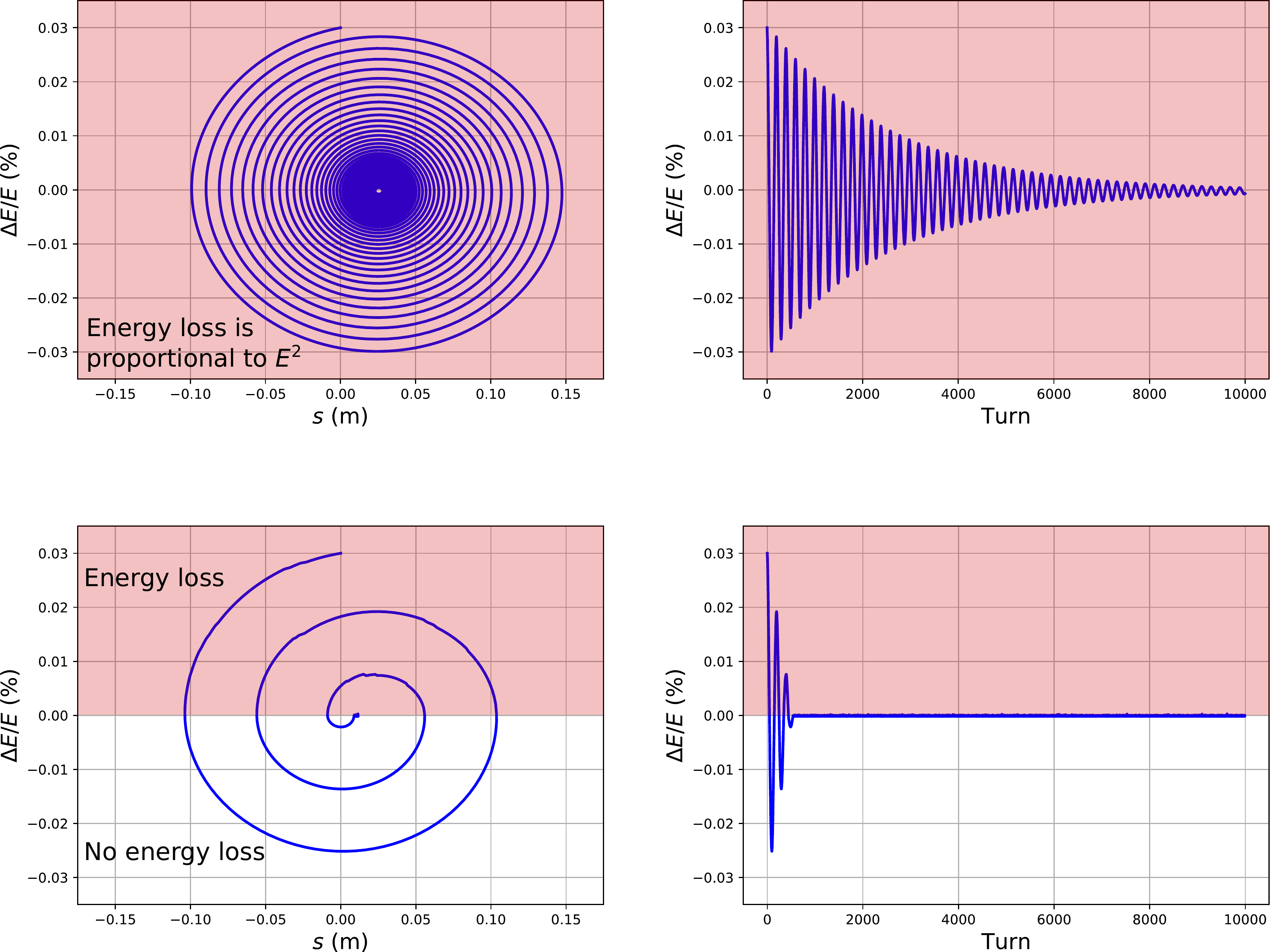}
	\caption{The evolution of the energy and the longitudinal 
	position of the ion, relative to their central values, 
	as a function of the turn number in the storage ring, for two
	regimes of the laser cooling. The top plots show the broad-band
	laser cooling \cite{BessonovKimPRL1996} using  the laser 
	frequency band which is large enough to excite all the ions,
	disrespectful of their energies. 
	The bottom plots show the regime of fast cooling
	\cite{Bessonov2008src} using the laser frequency band which has 
	a sharp cut-off,  positioned such that the ion absorbs the laser
	photon only if the ion energy is above its central value.}
    \label{fig:slow_vs_fast_cooling}
\end{figure}

A leap in the cooling speed can be achieved by
exciting intentionally only the fraction 
of the ions -- those which carry the energy 
larger than the central value of their energy spread.
This is possible because the width of the atomic transition 
and the laser-photon energy band  can be tuned to be 
much narrower than the energy spread in the ion beam.
In this case, the cooling time is comparable to the 
time which is necessary 
to radiate a fraction of the ion energy, which is 
of the order of the relative energy spread of the ion beam: $\sim 10^{-4}$--$10^{-5}$. 
The improvement of the cooling speed
is illustrated in Fig.~\ref{fig:slow_vs_fast_cooling}.

 \subsection{Transverse cooling}
 \label{Transverse-cooling}
 
 The transverse beam cooling accompany, in a natural way, the longitudinal one 
 because the effective friction force due 
 the emission of photons is directed opposite 
 to the ion momentum vector, while the RF-cavity restores only its 
 longitudinal component. This type of cooling is, however, too slow 
 to be used at the SPS flat-top energy\footnote{The resistive power 
 dissipated in the main SPS dipoles and the quadrupoles 
 at the top SPS energy is $44$~MW. The SPS beam can 
 be coasted at this energy (using the pulsed magnet operation mode) 
 over the time interval  which should be  shorter  than   
 $\tau_{coast}  \approx 15$ seconds. As a consequence, the beam 
 cooling phase  must be finalised  within this time interval. 
 A fallback solution would be  to cool the beam at 
 the energy below the value
 $E_{\rm beam} =  270 \times Z\,$GeV, at which the beam can be 
 permanently coasted, 
and resume the acceleration up to the LHC injection energy following 
the beam-cooling phase.}. 
 
\begin{figure}[!htpb]\centering
	\includegraphics[width=0.75\linewidth]{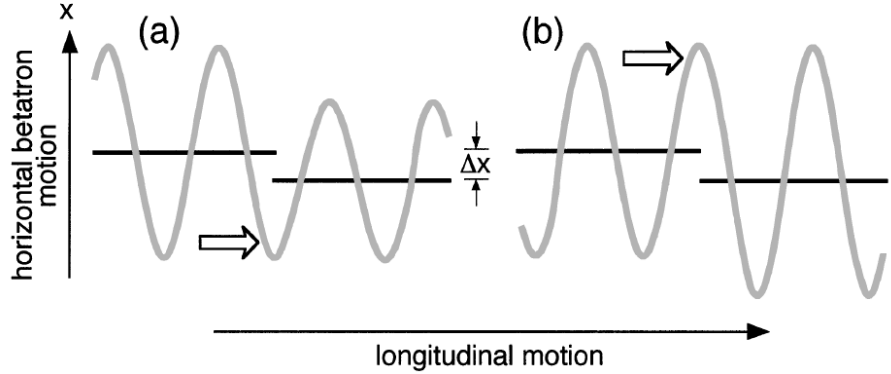}
	\caption{Horizontal betatron oscillations of a stored ion around the central orbit in a region with positive dispersion. The moment of photon emission and the corresponding change of the central orbit is indicated by the arrow. A reduction of the amplitude of the oscillation occurs when an ion radiates a photon at a negative ($x < 0$) phase of the betatron oscillation (a). If the photon is emitted at $x > 0$ (b), then the amplitude of
	the betatron oscillations is increased. The transverse cooling will occur if more photons
	are emitted at $x < 0$ than at $x > 0$. (Adapted from \cite{LauerPRL1998}.)
	}
    \label{fig:dispersive_coupling}
\end{figure}

 The cooling scheme which allows to shorten the cooling time below $15$ seconds 
 is based on the dispersive coupling of transverse 
 and longitudinal oscillations in a storage ring \cite{LauerPRL1998}. 
 The mechanism of the longitudinal--horizontal coupling through 
 dispersion is illustrated in Fig.~\ref{fig:dispersive_coupling}. 
 The horizontal betatron oscillations are first converted into energy 
 (synchrotron) oscillations and then the synchrotron oscillations are 
 suppressed  quickly using the fast longitudinal cooling method described 
 in the previous subsection. This scheme requires two different 
 lasers and two different photon--PSI interaction points. The focal point of 
 the first-laser beam is  shifted towards the negative horizontal position
 with respect to the ion beam centre (for a positive value of the dispersion function)
 by a value of $\Delta x$. 
 This laser has a broad frequency spectrum allowing to excite 
 the ions over the  full spread  of their energies. The focal 
 point of the second-laser beam is centred on the ion beam axis. Its frequency 
 band is  tuned  to excite only those of the ions which carry the energy  above 
 its central value.

 In order to suppress the vertical betatron oscillations, one needs 
 to couple them to the horizontal ones using the transverse 
 betatron coupling resonance. To achieve an efficient coupling, the frequency of the vertical betatron oscillations should be close enough to the frequency of the horizontal betatron oscillations.

 \subsection{Beam cooling R\&D in Gamma Factory PoP experiment}
 \label{PoP-experiment}

Evaluation of various techniques of the longitudinal and 
transverse cooling of atomic beams at ultra-relativistic 
energies will soon be  addressed in  the  forthcoming 
R\&D phase of the Gamma Factory project \cite{Krasny:2015ffb,Krasny:2018pqa} 
-- in its Proof-of-Principle (PoP) experiment \cite{GF-LoI}. 

This experiment plans to use the lithium-like lead beam, 
$\mathrm{Pb}^{79+}$. The lithium-like lead beam 
has been  chosen because it can be produced 
and accelerated at CERN with a minor change of the   
present operation mode of the fully stripped lead beams. 
The results 
of this experiment will then be  extrapolated to other 
ion species  specified by  $A$, $Z$ and the number 
of left electrons $N_e$. The extrapolation results 
will determine, together with other beam operation
aspects discussed in the next section, the most optimal 
ion-beam candidate for the high-luminosity 
LHC collisions of nuclear beams.

\section{Operation constraints}
\label{Operation}

\subsection{Parasitic beam burning}
 \label{Burning}
 
 The increase of the luminosity in collisions of nuclear partonic bunches  
 would be useless if a dominant fraction of the beam particles
 were ``burned-off'' in those of the  ultra-peripheral collisions which
 change the magnetic rigidity  of the beam particles.
 Such processes  give rise to   beam losses in the cold sections 
 of LHC rings and  represent a serious danger for the operation of the
 LHC superconducting magnets. 
 These processes have been extensively discussed in 
 \cite{Jowett:2016yfa}. 
 
 The electron--positron pair production 
 in which the electron is bound to one of the colliding nuclei has  
 already been a principal luminosity limiting factor -- already  
 for the low-luminosity $\mathrm{PbPb}$ collisions at the LHC. The  
 cross section of this process decreases  very quickly with the 
 decreasing charge of the nucleus, proportionally to $Z^7$. 
 While prohibiting the use of the $\mathrm{Pb}$ beam for  
 the high-luminosity mode of the LHC, it becomes negligible for 
 low-$Z$ ions.  
 
 For low-$Z$ ions, the process of photon-induced dissociation of 
 the nucleus with emission of a single neutron or a pair of neutrons --
 having a less strong $Z^3$-dependence -- is  the dominant 
 parasitic beam burning process. The neutron-loss cross section is larger 
 than the inelastic cross section for ions heavier than krypton 
-- for lighter ions it is less important, e.g.\ it represents 
 only $38\%$ of the total cross section  for the $\mathrm{CaCa}$ 
 collisions and $5\%$ for the $\mathrm{OO}$ collisions. 
 
 We thus conclude that the parasitic beam burning processes does not 
 represent an obstacle for the high-luminosity collisions of 
 partonic nuclear bunches provided that ions with $Z \leq 25 $
 are used in such a scheme.

\subsection{Photon fluxes revisited}
\label{photon-fluxes}

As discussed in Section~\ref{Photons}, in order to maximise the photon--photon
luminosity,  which is proportional to $Z^4$, 
large-$Z$ ions appear to be the best beam candidates. 
However, given the beam-burning constraint  
discussed above, the optimal $Z$ of the ion beam 
should represent a compromise between the photon--photon luminosity increase
due to the charge-coherence effects and the ``allowed'' luminosity limits  
driven by the  parasitic beam-burning processes.
Such a compromise was discussed already long time ago 
in \cite{Brandt:1994dd} for the case of two beam-burning collision points\footnote{One of the authors (MWK)  acknowledges  numerous 
discussions with Daniel Brandt on the use of light nuclei for the 
DESY, BNL and CERN research programmes at the time of designing  
the running modes for the  EIC project for DESY and later for BNL.}.

\begin{figure}[htpb]\centering
	\includegraphics[width=0.75\linewidth]{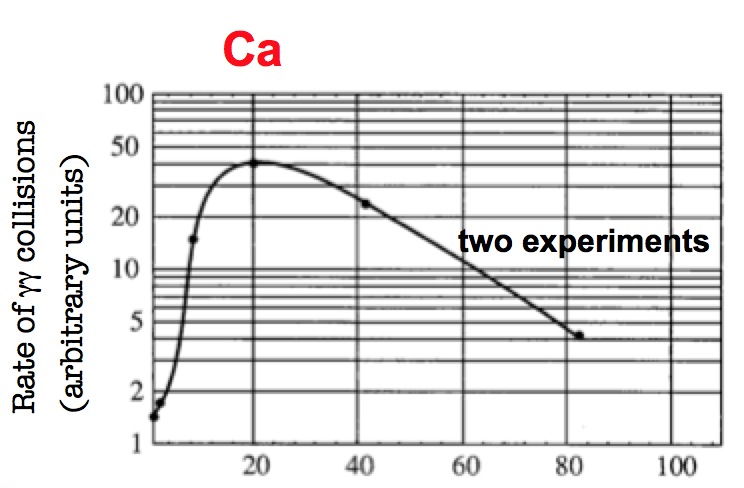}
	\caption{The $Z$-dependence of the effective rate of 
	the photon--photon collisions for the operation scenario discussed in \cite{Brandt:1994dd}.}
    \label{Ca_ph-ph}
    \vspace{-1.7cm}
    \hspace{7.9cm} {\large $Z$}
    \vspace{1.7cm}
\end{figure}

The highest effective  photon--photon luminosity, for the beam operation
model presented in \cite{Brandt:1994dd}, can be achieved for the $\mathrm{CaCa}$ collisions, see Fig.~\ref{Ca_ph-ph}.
It was found to be larger by a factor of $\approx 10$ w.r.t.\ the  $\mathrm{PbPb}$ collisions and by a factor of $\approx 30$ w.r.t.\ the $pp$ collisions. 
In reality, if no new forward ion detectors are designed and constructed
for the high-luminosity phase of the LHC operation, the increase  of the 
effective photon--photon luminosity  in the ``coherent-photon" kinematical 
domain is by far more significant. 
For the collisions of the calcium ions, the ``useful" photon--photon luminosity 
is expected to be by a factor of $\approx 4000$ larger than that for the
$pp$ collisions at the equivalent nucleon--nucleon 
luminosity\footnote{Photon--photon
collisions can be efficiently selected and measured only in the case of one beam particle collision per bunch crossing -- we remind that for the nucleon--nucleon luminosity of $2.5 \times 10 ^{34}\,$cm$^{-2}$s$^{-1}$, the event pile-up rate amounts to $\nu =150$
for the $pp$ collisions and $\nu =3.3$ for the $\mathrm{CaCa}$ collisions.}.

\subsection{Laser constraints}
\label{laser-constraints}

The beam-cooling process discussed in Section~\ref{Cooling} 
is based on resonant excitations
of the PSIs stored in the 
SPS ring by laser photons. The maximal energy, 
$E_{\rm max}^p$, of the SPS proton beam
and the wavelengths range  
of commercially available visible and 
near-infrared high-power lasers, 
$440\,\mathrm{nm} \leq \lambda \leq 3000\,\mathrm{nm}$, 
constrain the minimal number of
electrons which have to remain attached to their parent ions, $N_e$, 
at  the SPS phase of the beam acceleration  via the following condition:
\begin{eqnarray}
 440 < \frac{hcE_{\rm max}^p}{M_{\rm ion}Ry} 
 \times \frac{Z-N_e}{AZ^2} 
 \times (1+\cos\theta) 
 \times \frac{n_1^2(N_e)\,n_2^2}{n_2^2 - n_1^2(N_e)} <3000, 
\end{eqnarray}
where $M_{\rm ion}$ is the ion mass, $Ry$ is the Rydberg constant, 
$\theta$ is the collision angle
of the laser-photon beam and the PSI beam, $n_1(N_e)$
is the Bohr radial number of the ground state of the last 
(most loosely bound) of the $N_e$
electrons occupying the Pauli allowed levels and $n_2$ is 
the Bohr radial number of excited state.

The laser technology constraints do not restrict the $Z$ and $A$ range 
of ions which can be cooled at the SPS. It restricts, however, 
the number of electrons attached to their parent nuclei 
in the cooling phase of the beam. 
The number of electrons, $N_e$, must rise  
with the increasing $Z$ of the beam particles,
such that the increase of $n_1(N_e)$ compensates the $Z^2$-rise 
of the electron binding  energies. In practice, 
the beam of the hydrogen-like, $N_e=1$,
and the helium-like, $N_e=2$, oxygen ions can be cooled at the SPS with 
nearly head-on photon--ion collision angle. The minimal $N_e$ rises 
to $3$ for the calcium ions and to $11$ (fully filled 
the $n=1$ and $n=2$ atomic levels) for the krypton 
ions\footnote{These restrictions are important only for  
radial atomic excitations.
For the  excitation which do not change the Bohr radial 
number and which are driven by spin-orbit 
interactions, the $N_e=3$ limit can be kept even for the highest $Z$ ions, 
at the expense of significantly higher laser power requirement.
This option will be used in the Gamma Factory PoP experiment, 
and is not discussed in the present paper.}.

\subsection{Beam lifetime}
\label{SPS_loses}

For the SPS ring, where the  pressure of the residual gas
molecules stays  at the level of $10^{-8}\,$mbar (three order of magnitude higher 
than in the LEIR and the LHC),  there is a price to pay for
the increase of the number of electrons attached to their parent nucleus --
necessary for high-$Z$ ions -- the beam-lifetime decrease
below the SPS acceleration cycle length. 

The Gamma Factory group has performed 
dedicated machine studies  with partially stripped ion beams 
allowing to validate and calibrate the software tools used in calculation 
of the beam lifetime for arbitrary $Z$, $A$  and the number of left 
electrons, $N_{e}$ \cite{Winckler:2017aea}.

In 2017 the $^{129}\mathrm{Xe}^{39+}$ atomic beam was 
accelerated, stored in the SPS and studied at different 
flat-top energies \cite{Hirlaender:2018ipac,CERN_courier:2017,CERN_news:2018}.  
The analysis of the measured lifetime constrained 
the molecular composition of the residual gas in the SPS rings. 
In 2018, the  $^{208}\mathrm{Pb}^{81+}$ and $^{208}\mathrm{Pb}^{80+}$ 
beams were successfully injected to the SPS and accelerated to $270\,$GeV 
proton-equivalent energy. The observed lifetimes of the $^{208}\mathrm{Pb}^{80+}$ 
and $^{208}\mathrm{Pb}^{81+}$ beams of, respectively,  $350\pm 50$ and $600 \pm 30$
seconds  agreed with the predictions based on the calibrated molecular 
composition of the SPS vacuum  \cite{Schaumann:2019evk}.

The observed agreement allowed us to extrapolate the results of the SPS 
beam tests to arbitrary $Z$ and  $N_e$.
For high-$Z$ ions, such as xenon or lead, and high $N_e$, 
required to satisfy the laser technology 
constrains discussed in Section~\ref{laser-constraints}, the beam lifetime 
decreases significantly below the SPS cycle length for the LHC injection which is,
at present, approximately $20$ seconds.
For the very low-$Z$ ions, lighter than oxygen, the electrons become too loosely 
attached to their nuclei and no matter how many electrons 
are left attached, the SPS beam lifetime will always be shorter 
than the SPS cycle length\footnote{We are indebted to Slava Shevelko for his guidance and his calculations of the lifetimes of the partially stripped 
ions in the SPS.}. 

The ``sweet spot'' where the beam lifetime and the laser technology constraints 
are both satisfied is restricted to a very narrow region around $Z=20$.
For the present SPS vacuum conditions, 
the predicted lifetime of  $^{40}\mathrm{Ca}^{17+}$ beam 
is $16\pm 10$ seconds. If the SPS 
cycle length for the LHC injection is unchanged, the SPS vacuum quality 
would have to be improved by at least a factor of two to reduce the beam 
losses over the SPS stacking, acceleration and the cooling phases\footnote{The concentration of the vacuum molecules over the historical runs of the SPS
at the $p\bar p$ collider was lower by a factor of about $10$ with respect to the present values.}.

 \section{High-luminosity $\mathrm{CaCa}$ collision scheme}
 \label{Ca-Ca}
 
 The calcium (isoscalar nuclear) beam  
 satisfy the operation constraints, 
 discussed in the previous section and, in addition, has numerous merits 
 for the physics programme at the LHC, discussed in detail in Section~\ref{Merits}. 
 It is by far the most optimal candidate for the 
 high-luminosity operation of the LHC with a nuclear beam.

 In this section we present  a concrete scenario
 of producing, accelerating, cooling and colliding the calcium beams 
 at the LHC. The $\mathrm{Ca}$ beams have never been produced at CERN. 
 A more detailed scenario of their operation 
 in the CERN accelerator complex can only be worked out by the CERN 
 accelerator experts. In this section we  identify the most 
 critical points which would need to be addressed by such studies and 
 which are critical to prepare the calcium beam for the cooling phase in the SPS, and for its subsequent injection  to the LHC.
 We discuss also a concrete laser beam-cooling scheme of the $^{40}\mathrm{Ca}^{17+}$ beam and evaluate its expected performance.

 \subsection{Calcium source}
 \label{Ca-source}
 
The present CERN Linac~3 ion source is an $14.5\,$GHz ECR source optimised 
to produce the lead beams. The ECR source uses a long 
microwave heating pulse of $50\,$ms  which ionises lead atoms 
to the highest achievable charge states and subsequently forms 
$1\,$ms long ion-beam pulses. To match the pulse length to the Low Energy 
Ion Ring (LEIR) requirements 
only 1/5 of this pulse is accelerated in Linac~3 and injected to LEIR. 
This scheme can be used  for the operation of the calcium source. 

The required isoscalar calcium isotope of $_{20}^{40}\mathrm{Ca}$ has an abundance of $96.9\%$. The vapour pressure for calcium is higher than for lead, such that the ovens could be run $\sim 80\,$K cooler for the same vapour pressure \cite{Wolf:306831}. The change of the 
lead rod to the calcium rod is, in principle, easy but a setting-up time 
would be necessary for the overall optimisation of the source yield.  

The relative yields of principal-charge states after stripping 
of electrons at the exit of Linac~3 can be predicted by assuming the input energy 
of $4.2\,$MeV/u and the equilibrium charge state after stripping:
\begin{center}
	\begin{tabular}{|l|c|c|c|}
	\hline 
\rule{0pt}{12pt}Model	& Ca$^{16+}$ & Ca$^{17+}$ & Ca$^{18+}$ \\
	\hline 
\rule{0pt}{12pt}Baron {\it et al.} \cite{Baron:1993eg}	& 13\% & 38\% & 36\% \\
	\hline 
\rule{0pt}{12pt}Schiwietz {\it et al.} \cite{Schiwietz:2001bj}	& 20\% & 34\% & 27\% \\ 
	\hline 
\end{tabular} 

\end{center}

It is important to note that the requisite charge state
for the laser cooling at the SPS, Ca$^{17+}$, can already be
achieved at the exit of Linac~3  with the maximal 
efficiency. Therefore, contrary to the lead beam, no additional 
electron stripping is required in the PS-to-SPS transfer line.

\subsection{$\mathrm{Ca}^{17+}$ beam in LEIR, PS and SPS} 

LEIR receives long pulses of ions from Linac~3 and  transforms  
its long  pulses  into high-brilliance bunches 
by means of multi-turn injection, electron cooling and
accumulation. 
Important issues which would need to be addressed for  the LEIR 
and PS  acceleration phase of the Ca$^{17+}$ beam,  with respect
to the canonical operation of the Pb$^{54+}$ beam,  are the 
space-charge effects and the beam transfer limitations.

The {\it space-charge tune shift parameter}, 
$\Delta Q _{\rm SC}$,
rises proportionally -- at the fixed  bunch longitudinal and 
transverse emittance -- to $Z^2$ and $N_{ion} (Z,A)$,
and inversely proportionally to $A$  and $\gamma _L ^3$.
The  tune shift parameter for the Ca$^{17+}$ bunches 
of the same ion population as the Pb$^{54+}$ bunches is 
smaller\footnote{We assume here that LEIR will operate 
the Ca$^{17+}$ beam at the same magnetic rigidity as
the Pb$^{54+}$ beam.} by a factor of $8.7$. There is thus a room 
to increase the Ca$^{17+}$ bunch intensity by this factor 
while preserving exactly the same $\Delta Q _{\rm SC}$. For the 
requisite bunch intensity for high-luminosity
collisions, discussed in Section~\ref{bunch-intensity}, 
the space-charge tune shift parameter would have to be 
higher by $70\%$ with respect to its present value for 
the Pb$^{54+}$ bunches. The beam cooling at the SPS could easily 
compensate this rise, provided that such bunches will 
survive the LEIR, PS and SPS acceleration phases. 
This aspect deserves more detailed studies and tests. 

The magnetic  rigidity of the Ca$^{17+}$ beam is by a factor 
of $1.6$ smaller, at the equivalent beam momentum, than for the 
Pb$^{54+}$ beam. To preserve the beam optics in the 
transfer lines between the LEIR, PS and SPS rings,
the injection energies per nucleon  would have to be
increased while switching from the $\mathrm{Pb}$ to $\mathrm{Ca}$ beams
by a factor of $[(Z-N_e)_{\rm Ca} \times A_{\rm Pb} ]/ [(Z-N_e)_{\rm Pb} \times A_{\rm Ca}]$.
An alternative approach, based on the same beam injection 
energies and changed magnetic field in the transfer lines, 
can also be employed but at the cost of increase of the 
space-charge tune shift parameter. Dedicated studies of the 
beam transfer aspects are needed to optimise the transfer
of the $^{40}\mathrm{Ca}^{17+}$ ions in the injector transfer lines. 

 \subsection{Laser cooling at the SPS}
 \label{Ca-cooling}
 
     \subsubsection{Laser--Ca-beam collision parameters}
     \label{laser-FP}

The Ca$^{17+}$-beam cooling, proposed below, uses the atomic $2s$--$3p$ transition. 
The energy of the $2s$--$3p$ transition is $661.89\,$eV. For the 
optimal phase of  cooling at the flat-top of the SPS 
acceleration cycle (just before the extraction to the LHC), 
the Lorentz factor of the beam should be $\gamma _L=205$. 
The incoming laser-photon energy is $2\gamma _L$ times lower than 
the atomic excitation energy in its rest frame 
-- at the top SPS energy it should be equal to $1.6\,$eV. The corresponding wavelength 
of the laser photons is $768\,$nm  for the nearly head-on collision of 
the laser beam with the ion beam.
The full set of parameters for the Ca$^{17+}$-beam cooling 
configuration in the SPS is summarised in Table~\ref{tab:CoolingParameters}.

\begin{table}[!htpb]\centering
\caption{Parameters of the calcium-beam cooling configuration in the SPS.}
	\begin{tabular}{lr}\toprule\toprule
		Ion beam  & ${}^{40}\rm{Ca}^{17+}$ \\
		\hline
		$m$ -- ion mass & $37.21$ GeV/c$^2$ \\
		$E$ -- mean energy & $7.65$ TeV \\
		$\gamma _L=E/mc^2$-- mean Lorentz relativistic factor & $205.62$ \\
		$N$ -- number ions per bunch & \num{4e9} \\
		$\sigma_E/E$ -- RMS relative energy spread & \num{2e-4}  \\
		$\epsilon_n$ -- normalised transverse emittance & \SI{1.5}{mm.mrad}\\
		$\sigma_x$ -- RMS transverse size & \SI{0.80}{mm}\\
		$\sigma_y$ -- RMS transverse size & \SI{0.57}{mm}\\
		$\sigma_z$ -- RMS bunch length & \SI{10}{cm}\\
		Dispersion function & $2.44$ m\\
		\hline
		Laser & pulsed Ti:Sa (20MHz)\\
		\hline
		$\lambda$ -- wavelength ($\hbar\omega$ -- photon energy) & $768$ nm ($1.6$ eV) \\
		$\sigma_{\lambda}/\lambda$ -- RMS relative band spread & \num{2e-4}  \\
		$U$ -- single pulse energy at IP & \SI{2}{mJ} \\
		$\sigma_L$ -- RMS transverse intensity distribution at IP ($\sigma_L=w_L/2$) & \SI{0.56}{mm}\\
		$\sigma_t$ -- RMS pulse duration & \SI{2.04}{ps} \\
		$\theta_L$ -- collision angle & $1.3$ deg \\
		\hline
 		 Atomic transition of ${}^{40}\rm{Ca}^{17+}$ & $2s \rightarrow 3p$\\
		\hline
		$\hbar\omega'_0 $ -- resonance energy & $661.89$ eV\\
		$\tau'$ -- mean lifetime of spontaneous emission & $0.4279$ ps\\
		$\hbar\omega_{1}^{\rm max}$ -- maximum emitted photon energy & $271$ keV\\
		\bottomrule\bottomrule
	\end{tabular}	
\label{tab:CoolingParameters}	
\end{table}
 
 The ion-beam parameters assume that the beam-cooling interaction
 point (IP) will be placed at the half-cell 627 of the SPS, in the place of the 
 planned Gamma Factory PoP experiment \cite{GF-LoI}. 
 A commercial {\sf Titanium:Sapphire} mode-lock laser 
 oscillator\footnote{We are indebted to Kevin Cassou and Aurelien Martens 
 for the numerous discussions and sharing with us their knowledge
 on the available laser technology.} ,
 which provides excellent phase noise stability, fulfils the
 requisite requirements for the photon source. 
 The effective photon flux is  
 increased to its requisite value by using a high-gain Fabry--P\'erot resonator 
 with a typical enhancement factor of $10^4$~\cite{GF-LoI,Borzsonyi:13}.

\subsubsection{Simulation results}
\label{Simulation-results}
     
Monte Carlo turn-by-turn simulations of ion-bunch dynamics 
during the cooling process have been performed 
using the full-turn 6D transport matrix for the SPS lattice
\cite{GF-Python_Ca_in_SPS_transv}. 

\begin{figure}[htpb]\centering
	\includegraphics[width=0.75\linewidth]{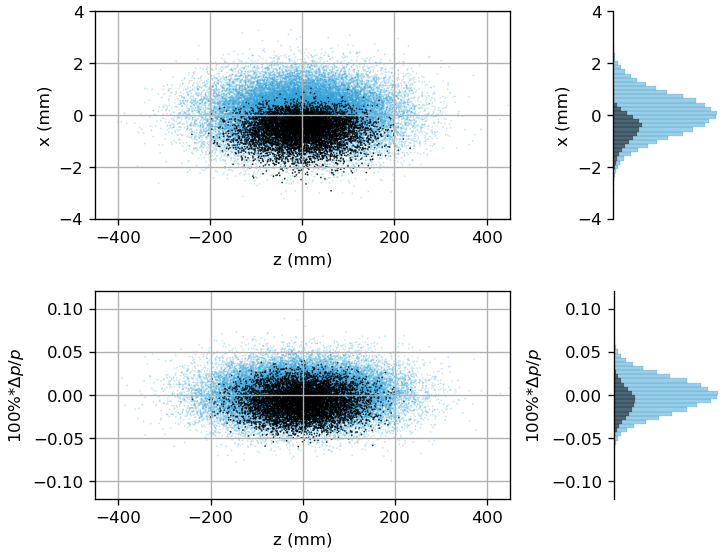}
	\caption{Distributions of the positions and momenta of 
	the ions interacting  
	with the pulse of the first laser. Excited 
	ions are shown as black dots while non-excited ions are shown as 
	blue dots. The shift of the laser pulse by 
	$-1.4\,$mm  provides an optimal coupling of  horizontal betatron 
	oscillations to synchrotron oscillations, as explained in
	Fig.~\ref{fig:dispersive_coupling}. About $17\%$ of all ions
	are excited in each bunch crossing.}
    \label{fig:z-x_excited_ions}
\end{figure}
     
In order to couple the horizontal betatron oscillations to 
the vertical ones, the betatron tunes in the uncoupled case 
are set to the same value: $\nu_x=\nu_y=26.13$ (while the 
design values are only slightly different: $\nu_x = 26.13$,
$\nu_y = 26.18$). In such a case, the transverse coupling is introduced 
with a single $1\,$m long skew-quadrupole with a strength which 
is approximately $10$ times lower than the typical SPS quadrupole 
strength. The resulting width of the coupling resonance 
(tune separation) is $0.0078$. This means that the vertical 
betatron oscillations are transferred into the horizontal 
oscillations in about $100$ turns.

In order to couple the transverse oscillations 
to the synchrotron oscillations, the focal point of the 
first laser beam, of the $2\,$mJ pulse power,
is shifted by $1.4\,$mm in the negative horizontal 
direction with respect to the ion-beam axis. 
The resulting distributions of the positions 
and momenta of the single-bunch ions 
which are excited by a single laser pulse 
are  shown in Fig.~\ref{fig:z-x_excited_ions}.
The number of the excited ions at $\Delta p<0$, as shown in
Fig.~\ref{fig:z-x_excited_ions}, is slightly larger than 
the number of the ions excited at $\Delta p>0$. 
This leads to the longitudinal emittance blow-up, 
unless it is ``corrected" with the second laser. 
This additional laser allows to counteract the longitudinal emittance 
blow-up induced by the transverse cooling.
The focal point of its photon pulses is 
aligned, in the transverse plane, with the centre of the ion beam spot.

\begin{figure}[!htpb]\centering
	\includegraphics[width=0.75\linewidth]{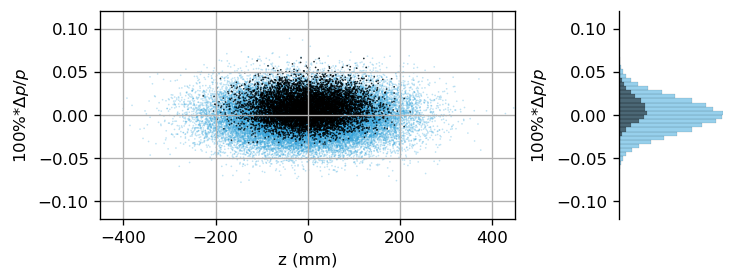}
	\caption{Distribution of the momentum and longitudinal 
	positions of the ions interacting with the photon-pulse of the second
	laser. Excited ions are shown as black dots while non-excited  ions are shown as blue dots. The laser pulse focal point is aligned 
	with the ion beam centre but its frequency band is shifted to excite the higher-momentum ions, as explained in Fig.~\ref{fig:slow_vs_fast_cooling}.}
    \label{fig:excited_ions_2nd_laser}
\end{figure}

The frequency-band  of the second laser is tuned  to 
excite the ions predominantly  in the upper part of their  momentum distribution, 
as shown in Fig.~\ref{fig:excited_ions_2nd_laser}. Its pulse power 
is two times smaller than that of the first one. 
This configuration assures stable transverse cooling while 
preserving the longitudinal bunch size. If  
the power of the second-laser pulse were increased, 
the ion-bunch length would be decreased.  
This regime is also interesting to be considered. However,
it would require supplementary studies of the longitudinal stability 
of short ion bunches in the SPS. 
By keeping the longitudinal bunch size unchanged over the 
transverse beam cooling phase, the bunches are protected against 
longitudinal instabilities, e.g.\ the microwave instability,
and the intra-beam scattering rate of the beam particles is reduced.
The transverse stability 
\cite{Bartosik:2014ria, Zannini:2015eqc,Carla:2019tto}
of the low-emittance ion beam 
in the SPS and LHC deserves complementary studies. They 
should involve investigations of the transverse mode coupling 
and head--tail instabilities.

\begin{figure}[!htpb]\centering
	\includegraphics[width=0.75\linewidth]{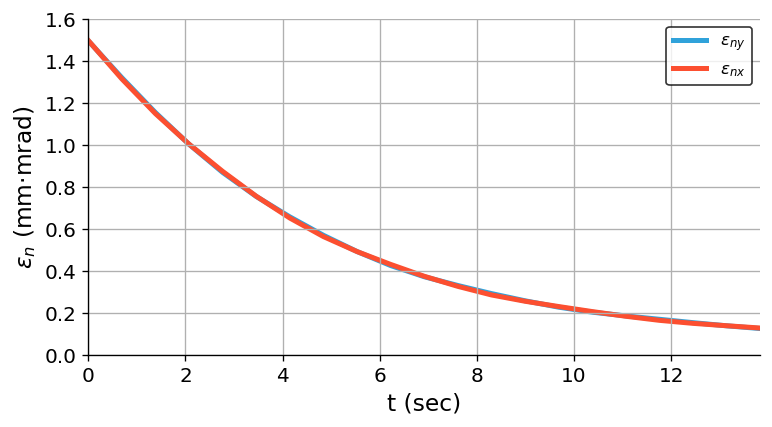}
	\caption{Transverse cooling speed: the time-evolution curves of the vertical and horizontal
	emittances are overlapping each other -- they 
	are precisely equal when the betatron tunes are on the coupling resonance.}
    \label{fig:emittance_vs_time}
\end{figure}

The evolution of the transverse beam emittance in the proposed 
scheme is shown 
in Fig.~\ref{fig:emittance_vs_time}. A factor of $3$ reduction of the transverse emittance can be  achieved in 5 seconds and a factor of $5$ in 8 seconds.

\subsection{$\mathrm{Ca}^{20+}$ beam in LHC}
\label{Stability}

The three electrons remaining attached to the $\mathrm{Ca}$ nucleus 
over the beam cooling process in the SPS storage ring 
have to be stripped in the 
TI2 and TI8 transfer lines before accelerating the ion bunches in the LHC. 
Dedicated studies will be required  to 
optimise the stripper material and thickness, such that 
the maximal Ca$^{20+}$ transmission efficiency is associated 
with a negligible increase of the beam momentum dispersion, 
caused by the ion energy loss, and with the minimal emittance
blow up, caused by multiple scattering of the ions in the
stripper material. Optimisation of the position of 
the stripper may be needed for the present Twiss function 
of the transfer line (the stripper-induced emittance growth is
minimal for placing the stripper at the the minimal beta point). 
Note that the above constraints are  less critical 
than the corresponding constraints for the stripping of the 
Pb$^{54+}$ ions in the PS to SPS TT2 line \cite{Arduini:1996up}
because of: (1) significantly larger (by a factor of $30$) ion-beam
energy at the stripper position, (2) a smaller number (by a
factor of $10$) of electrons which need to be stripped off and
(3) significantly smaller (reduced by a factor of $Z^2_{\rm Ca}/Z^2_{\rm Pb}$) 
binding energies of electrons. 
The latter two factors allow to reduce considerably 
the stripper thickness. 
According to  initial estimates, the emittance blow-up in the stripper 
can be kept smaller than $\Delta \epsilon _n = 0.1\,$mm$\,$mrad 
and should not contribute to a sizeable emittance blow-up.

Another, more important, effect which would need to be studied
is the effect of the blow-up of the beam transverse emittance
over the phase of stacking the SPS batches into the SPS ring and
over the phase of the beam acceleration. The emittance blow-up is
caused by the intra-beam scattering of the beam particles.
Several techniques may be applied to counterbalance the
corresponding emittance growth and, as a fallback solution,  
the stochastic cooling at the top beam-collision energy 
may need to be employed  to mitigate the intra-beam scattering 
driven emittance growth.

Let us note that at the Relativistic Heavy Ion Collider (RHIC),
the stochastic cooling of the bunched
gold and uranium beams has been very successful 
\cite{Blaskiewicz:2007zza,Blaskiewicz:2008zz,Schaumann:2014lba}. 
The emittance growth was substantially reduced, improving 
the luminosity lifetime.  For the uranium operation
it was possible even to increase the instantaneous 
luminosity by more than a factor
of $3$ over its initial value. Such a  method could also 
be exploited for the LHC ion beams \cite{Schaumann:2014lba}. 
Another fallback option would be to implement the optical
stochastic cooling method, specially designed for 
the high-brightness beams \cite{Lebedev:2014cha}. 
Both methods could help in 
preserving the initial, LHC-injection phase, beam transverse 
emittance.

The optical stochastic cooling may turn to be necessary not only 
to reduce the intra-beam scattering emittance growth but also 
to be able run the collisions of the low-emittance beams with
a  high value of the beam--beam parameter.
For the equivalent and maximal partonic luminosity, the beam--beam parameter 
in the $\mathrm{CaCa}$ collisions at the LHC would be  significantly higher 
than that for the $pp$ collisions. It would reach  the value of  
$\xi_{x,y} \approx  0.1$,  which is by a factor of $3$  
larger than the one at the Tevatron collider. 
Running such a collisions scheme is anything but easy.
It may require beam cooling and/or the use of electron lenses\footnote{We are indebted to Frank Zimmermann for drawing our attention to this potentially important luminosity-limiting factor.}.

Finally, an additional aspect,  which will require dedicated
studies,  is the beam collimation for the fully stripped 
calcium beam in the LHC. The LHC beam collimation system was
designed for the proton beam and the losses
of the beam particle that are specific for the nuclear 
beams, such as the neutron losses -- leading to a change of the
magnetic rigidity of the beam particles -- would need to be
evaluated. The reduced emittance of the beam may be of help for
the collimation of the high-intensity calcium beam in the LHC
rings.

 \subsection{Luminosity of $\mathrm{CaCa}$ collisions}
 \label{scenario}

The reduction of the transverse beam emittance of the calcium
beam, at the flat-top SPS energy, to the value of $\epsilon _n = 0.3\,$mm$\,$mrad can be achieved,
according to our simulations, within the cooling time of $8$ seconds.

Let us assume that the normalised transverse emittance 
can be preserved  over the time of the LHC fill and ramp or 
that the emittance growth in the LHC can be (if necessary)
compensated by the stochastic cooling at the top beam energy. According to  the formula~(\ref{lumi}), by using the high-luminosity LHC $pp$-mode parameters 
and the bunch population of the calcium beam taken from  
\cite{Citron:2018lsq}, the predicted peak nucleon--nucleon
luminosity (or equivalently the partonic luminosity) in the
$\mathrm{CaCa}$ collision mode with the $50\,$ns bunch interval
is expected to be
\begin{equation}
L_{NN}= 4.2 \times 10^{34} \mathrm{s}^{-1}\mathrm{cm}^{-2}.
\label{eq:LNN}    
\end{equation}
The expected number of collisions per bunch crossing is 
$\nu_{\rm CaCa} = 5.5$.
Note that for the same partonic luminosity, the number of 
collisions in the $pp$ running mode would be $ \nu_{pp} = 702$ --
too large to be accepted for efficient operation 
of the ATLAS and CMS detectors. For the $\mathrm{CaCa}$
collisions,  the number of produced particles, assuming the
``wounded-nucleon scaling'' discussed in Section~\ref{Pileup},
is expected to be lower at this luminosity  than those  
for the $50\,$ns $pp$ running mode at the levelled luminosity of
$L_{pp}= 2.5 \times 10^{34}$ s$^{-1}$cm$^{-2}$, 
and for the $25\,$ns mode with the twice higher levelled luminosity. 

The expected  beam transverse emittance growth at such a luminosity 
is of the order of 1--2 hours. The expected  optical stochastic 
cooling time for the $\mathrm{Ca}$ beam at the top energy 
(in its simplest scheme of passive cooling 
with no optical amplifier) is $1.5$ hours. The optical stochastic cooling 
should thus allow to reach the equilibrium between the beam cooling and
beam heating processes at the above luminosity\footnote{We are indebted to
Valeri Lebedev for his  calculations of the optical 
stochastic cooling time for the calcium-ion beam.}.

\section{Conclusions and outlook}
\label{conclusions}

We have argued  that nuclear beams, 
and in particular isoscalar nuclear beams, 
have numerous  advantages with respect to the proton beams for  
the high-luminosity phase of the LHC operation. 
They allow to make the full use of the isospin symmetry of the strong interactions 
in  constraining  the flavour and momentum distributions 
of their partonic degrees of freedom.
As a consequence, the analysis of the SM EW and BSM 
processes  becomes insensitive  to the  
limited precision of the LHC-external constraints,  
which will always remain essential for the interpretation 
of the numerous $pp$ collision measurements. 
The nuclear beams  can serve as  analysers (targets)
to study the propagation asymmetries of the longitudinally 
and transversely polarised $W$ and $Z$ bosons in vacuum and matter.  
They can generate effective photon beams of unprecedented intensity,
unreachable with the proton beams, 
opening the path to studies of the Higgs-boson 
production in photon--photon collisions. Finally, they 
reduce the pile-up background in the high-luminosity 
phase of LHC operation while preserving the 
same  partonic luminosity or, conversely,  increase the HL-LHC
partonic luminosity at a fixed soft-particle pile-up level.

We have proposed a new operation scheme of nuclear beams
which includes a significant reduction of  
their transverse emittance by laser cooling.  
This scheme creates a possibility to reach, for low-$Z$ nuclear beams,    
a comparable partonic luminosity to that foreseen  
for the high-luminosity phase of $pp$ collisions.

A concrete scenario for the calcium beams is  presented and evaluated.  
In this scenario, the Ca$^{17+}$ ion beam coming from the ion source 
is cooled at the SPS flat-top energy and its transverse emittance 
is reduced over the cooling time of $15$ seconds to the value of
$\epsilon _n = 0.3\,$mm$\,$mrad. Assuming that the transverse 
emittance of such a beam is preserved over the phase of stacking 
of the SPS batches in the LHC and over its acceleration phase,  
the expected peak nucleon--nucleon (partonic) luminosity for the  
$\mathrm{CaCa}$ collisions is $L_{NN}= 4.2 \times 10^{34}\,$s$^{-1}$cm$^{-2}$.
This value is higher 
than the levelled value for the high-luminosity $pp$ collisions
with $50\,$ns bunch spacing and similar to that for the $25\,$ns mode. 
This scenario deserves further studies. The most important aspects  
of such remaining studies  have been identified and discussed. 

The cold light-ion beams are, in our view, the best candidates 
to achieve the highest luminosity at the future high-energy hadron 
colliders, e.g.\ the FCC-hh. The cooling efficiency in the LHC 
(if used as a injector to the FCC-hh) will be higher w.r.t.\ the SPS.
The LHC vacuum is by a factor of $1000$ better than that of the SPS,
and, most importantly, the blow up of the transverse beam 
emittance over the time  interval  between the beam-cooling phase 
and the beam-collision phase is significantly reduced due to increase 
of the Lorentz $\gamma_L$-factor of the beam. This is another reason 
why the proposed scheme should be considered seriously,  
elaborated in more detail and allowed to be proven in the dedicated 
runs of the Gamma Factory Proof-of-Principle experiment at the SPS.

\section*{Acknowledgements}

We would like to thank all the members of the Gamma
Factory group for numerous discussions and encouragement.
Particular thanks to:
Reyes Allemany-Fernandez,
Hannes Bartosik, 
Evgeny Bessonov,
Roderick Bruce,
Dima Budker,
Kevin Cassou,
Camilla Curatolo,
Yann Dutheil,
Brennan Goddard,
John Jowett, 
Detlef Kuchler,
Mike Lamont, 
Valeri Lebedev,
Thibaud Lefevre,
Aurelien Martens, 
Michaela Schaumann,
Richard Scrivens,
Slava Shevelko, 
Vladi\-mir Shiltsev 
and Frank Zimmermann. 

A.~Petrenko acknowledges a personal grant from the Foundation for the Advancement of Theoretical Physics and Mathematics ``BASIS''.

\newpage

\begin{appendices}

\appendix

\section{Parton distribution functions for protons and nuclei}
\label{annex:PDFa}

Parton distribution functions (PDFs) were introduced by Feynman in 1969 to explain 
the Bjorken scaling in deep inelastic scattering data. They have interpretation
of probability distributions of finding a given parton within a hadron or a nucleus 
with a longitudinal momentum fraction $x$ of its parent at a resolution scale $Q^2$. 
Within QCD/QED the partons are interpreted as quarks, gluons and photons. 
According to the QCD factorisation theorem for inclusive hard-scattering processes, 
PDFs are universal distributions containing long-distance (non-perturbative) structure
of hadrons/nuclei. They are key ingredients for phenomenology of high-energy 
hadron/nucleus collisions.

Since PDFs are non-perturbative objects, they cannot be calculated from first principles
using perturbative QCD methods\footnote{There are attempts to obtain PDFs using
non-perturbative techniques of lattice QCD, but such PDFs are still far from practical
applications for the HEP phenomenology, 
see e.g.~\cite{Liu:2018uuj} and references therein.}. 
Instead, they are usually parametrised at some initial scale $Q_0^2$ with the
help of a set of chosen parameter-dependent functions. Then, they are evolved
to any scale $Q^2>Q_0^2$ using the DGLAP equations \cite{DGLAP} at a given order of
perturbative QCD (LO, NLO, NNLO) and fitted to experimental data for the used
parameters/functions.
In this way an appropriate PDFs parametrisation is obtained and it is usually supplied
with some uncertainties corresponding to $1\sigma$ errors of the respective likelihood analysis.
Fits are performed for various kinds of data from deep inelastic lepton--nucleon/nucleus
scattering as well as hadron--hadron, hadron--nucleus and nucleus--nucleus collision
experiments in order to have as good as possible parton-flavour sensitivity and 
$x$-range coverage.
Currently, there are several groups that provide independent PDFs parametrisations
using different assumptions and/or analysis methods -- for a recent review see
e.g.\ Ref.~\cite{Ethier:2020way}.

\begin{table}[!htpb]\centering
{\small
\begin{tabular}{lcccc}
\toprule
\toprule
  Proton PDFs 
& {\sf CT18}
& {\sf MMHT14}
& {\sf NNPDF3.1}
& {\sf ABMP16}
\\
\midrule
  Perturbative order      
& NLO, NNLO  
& LO, NLO, NNLO 
& LO, NLO, NNLO 
& NLO, NNLO 
\\ 
  Heavy-quark scheme      
& S-ACOT     
& optimal-TR    
& FONLL   
& FFN    
\\
  Value of $\alpha_s(m_Z)$ 
& 0.118      
& 0.118         
& 0.118   
& fitted    
\\
  Input scale $Q_0$
& 1.30 GeV
& 1.00 GeV
& 1.65 GeV
& 3.00 GeV
\\
\midrule
  Fixed Target DIS    
& \checkmark 
& \checkmark 
& \checkmark 
& \checkmark 
\\
  Collider DIS        
& \checkmark 
& \checkmark 
& \checkmark 
& \checkmark 
\\
  Fixed Target SIDIS  
&            
&            
&            
& \checkmark 
\\
  Fixed Target DY     
& \checkmark 
& \checkmark 
& \checkmark 
& \checkmark 
\\
  Collider DY         
& \checkmark 
& \checkmark 
& \checkmark 
& \checkmark           
\\
  Jet production      
& \checkmark 
& \checkmark 
& \checkmark 
& \checkmark            
\\
  Top production      
& $t\bar{t}$ tot., diff.
& $t\bar{t}$ tot.
& $t\bar{t}$ tot., diff.
& $t\bar{t}$, single-$t$, tot.           
\\
  \midrule  
  Independent PDFs 
& 6
& 7
& 8
& 6
\\
  Parametrisation
& Bernstein pol. 
& Chebyshev pol.
& neural network 
& simple pol.
\\
  Free parameters
& 29
& 37
& 296
& 25
\\
  Statistical treatment     
& Hessian
& Hessian
& Monte Carlo
& Hessian 
\\
  Tolerance
& $\Delta\chi^2=100$
& $\Delta\chi^2$ dynamical
& n/a
& $\Delta\chi^2=1$         
\\
\bottomrule
\bottomrule
\end{tabular}
}
\caption{Theoretical, experimental and methodological features of some recent proton PDF
              sets, from Ref.~\cite{Ethier:2020way}.} 
\label{tab:pPDFs}
\end{table}

The proton PDFs are known to a much better precision than the nuclear ones as they have
been the main focus of phenomenology for the lepton--proton and proton--(anti)proton 
collider experiments over the last 50 years. 
In Table~\ref{tab:pPDFs} we summarise theoretical, experimental and methodological
features of the recent proton PDF parametrisation sets:
{\sf CT18}~\cite{Hou:2019efy}, {\sf MMHT14}~\cite{Harland-Lang:2014zoa},
{\sf NNPDF3.1}~\cite{Ball:2017nwa} and {\sf ABMP16}~\cite{Alekhin:2017kpj,Alekhin:2018pai};
based on Ref.~\cite{Ethier:2020way}. 
The {\sf MMHT14} and  {\sf NNPDF3.1} sets have been updated since their first publications,
taking into account new experimental data and improving theoretical frameworks;
for more details see e.g.\ Ref.~\cite{Ethier:2020way}. 
Other proton PDF parametrisations currently available are 
{\sf JAM19}~\cite{Sato:2019yez}, {\sf JR14}~\cite{Jimenez-Delgado:2014twa}, 
{\sf CJ15}~\cite{Accardi:2016qay} and {\sf HERAPDF2.0}~\cite{Abramowicz:2015mha},
although they are based on smaller sets of measurements than the ones listed 
in Table~\ref{tab:pPDFs}.

\begin{table}[!htpb]\centering
{
\begin{tabular}{lcccc}
\toprule
\toprule

Nuclear PDFs 
& {\sf nCTEQ15}
& {\sf EPPS16}
& {\sf nNNPDF1.0}
& {\sf TUJU19} 
\\
\midrule
  Perturbative order
& NLO
& NLO
& NLO, NNLO
& NLO, NNLO
\\
  Heavy quark scheme
& ACOT
& S-ACOT
& FONLL
& ZM-VFN
\\
  Value of $\alpha_s(m_Z)$
& 0.118
& 0.118
& 0.118
& 0.118
\\
  Input scale $Q_0$
& 1.30 GeV
& 1.30 GeV
& 1.00 GeV
& 1.30 GeV
\\
\midrule
  Fixed Target DIS
& \checkmark
& \checkmark
& (w/o $\nu$-DIS)
& \checkmark
\\
  Fixed Target DY
& \checkmark
& \checkmark
&
&
\\
  Colider DY
&
& \checkmark
& 
&
\\
  Jet and had. prod.
& ($\pi^0$ only)
& ($\pi^0$, dijet)
& 
&
\\
\midrule
  Independent PDFs
& 6
& 6
& 3
& 6
\\
  Parametrisation
& simple pol.
& simple pol.
& neural network
& simple pol.
\\
  Free parameters
& 16
& 20
& 178
& 16 
\\
  Statistical treatment
& Hessian
& Hessian
& Monte Carlo
& Hessian
\\
  Tolerance
& $\Delta\chi^2=35$
& $\Delta\chi^2=52$
& n/a
& $\Delta\chi^2=50$
\\
\bottomrule
\bottomrule
\end{tabular}

}
\caption{Theoretical, experimental and methodological features of the recent nuclear PDF
              sets, from Ref.~\cite{Ethier:2020way}.} 
\label{tab:nPDFs}
\end{table}

The need for precise nuclear PDFs has
been driven recently mainly by heavy-ion collisions at the LHC.
A summary of the theoretical, experimental and methodological
features of the recent nuclear PDF parametrisation sets
{\sf nCTEQ15}~\cite{Kovarik:2015cma}, {\sf EPPS16}~\cite{Eskola:2016oht},
{\sf nNNPDF1.0}~\cite{AbdulKhalek:2019mzd} and {\sf TUJU19}~\cite{Walt:2019slu}
is presented in Table~\ref{tab:nPDFs}; based on Ref.~\cite{Ethier:2020way}. 
The {\sf nCTEQ15} and {\sf EPPS16} sets have been updated since their original publications
with some new experimental data, for more details see e.g.\ Ref.~\cite{Ethier:2020way}.
The nuclear PDFs are usually provided in terms of the correction factors
$R_i^A(x,Q^2)$ with respect to the proton PDFs, as given by Eqs.~(\ref{eq:RAq})
and (\ref{eq:RAg}) when integrated over $k_t$. The present precision of nuclear 
PDFs is significantly worse that the proton PDFs. 
Most of the above proton and nuclear PDF sets are publicly available through
the {\sf LHAPDF} library \cite{Buckley:2014ana}.

The advantage of the PDFs is that they are universal and can be combined
with theoretical perturbative calculations for any hard parton-level process 
in high-energy lepton--hadron/nucleus and hadron/nucleus--hadron/nucleus collisions. 
For many Standard Model processes at the LHC, these theoretical predictions are given
at the next-to-next-to-leading (NNLO) accuracy in the strong coupling constant $\alpha_s$.
These PDFs have, however, an important drawback -- they are integrated over the transverse
degrees of freedom of the respective partons, so they do not provide any information
on the distributions of their  transverse momenta $k_t$. 
Many observables at the LHC are sensitive to the transverse momenta 
of initial partons, in particular to their intrinsic (non-perturbative) $k_t$.
This is particularly important for precision measurements of the Standard Model
parameters, such as the $W$-boson mass and the weak-mixing angle.
This is the reason why, in Section~\ref{Partonic_beams} we have not referred 
to these ``collinear'' PDFs but used a more general notion of partonic distributions 
that depend on both $x$ and $k_t$.

In general, transverse-momentum-dependent (TMD), or ``unintegrated'', PDFs, 
contrary to the above collinear PFDs, are not process independent, see e.g.~\cite{Avsar:2011tz}.
However, they can be defined unambiguously for specific high-energy scattering processes,
such as semi-inclusive deep-inelastic scattering (SIDIS), the Drell--Yann process, 
the Higgs-boson production in $pp$ collisions, 
see e.g.\ Refs.~\cite{Angeles-Martinez:2015sea,Bacchetta:2019sam,Boer:2011kf}. 
Some parametrisation sets of this kind of TMD PDFs for protons and lead nuclei are 
publicly available through the {\sf TMDlib} library \cite{Hautmann:2014kza,Blanco:2019qbm}.

\newpage


\section{Photon absorption and emission by ultra-relativistic partially stripped ions}
\label{annex:kinematics}

\subsection{Kinematics}

\begin{figure}[htpb]\centering
\includegraphics[width=0.95\linewidth]{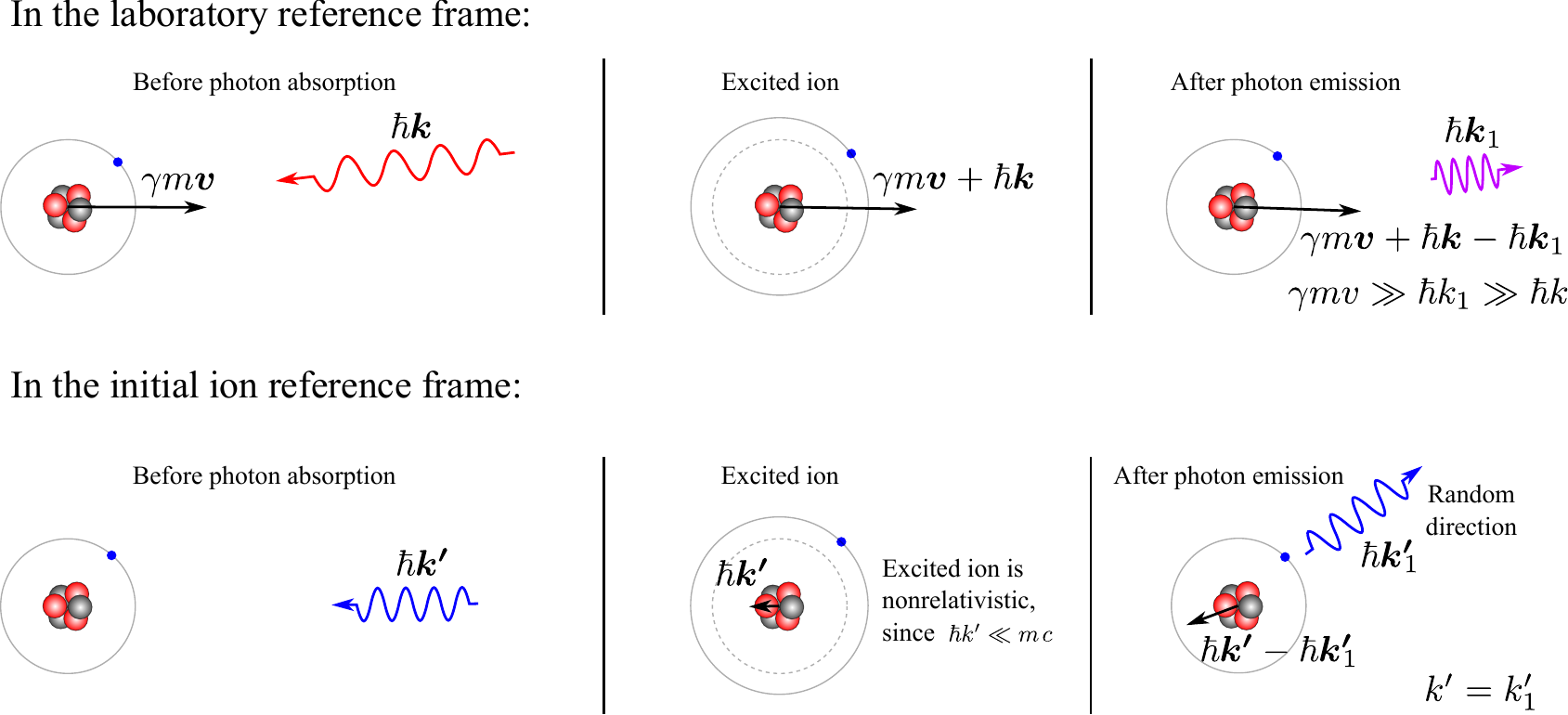}
 	\caption{The process of photon scattering in the laboratory 
 	and ion-rest reference frames.}
     \label{fig:lab_frame_vs_ions_frame}
\end{figure}

Interaction of photons with ultra-relativistic 
partially stripped ions is the key phenomenon which 
drives the beam-cooling process. 
Let us consider counter-propagating beams of laser
photons and ions.
Due to the relativistic Doppler shift, the energy of the laser photons
is  boosted in the ion rest frame  proportionally to the 
Lorenz $\gamma_L$-factor of the ion beam.
If tuned to atomic  transition energy of the ions,   
the laser photons are resonantly absorbed with a large
(gigabarn-level) cross section.
Excited ions eventually decay into their 
ground state by emitting photons. This process is shown in Fig.~\ref{fig:lab_frame_vs_ions_frame}.
For the ultra-relativistic ions, this absorption and emission 
scheme enables a conversion of infra-red, visible or near-ultraviolet
photons into X-ray or $\gamma$-ray photons.

\begin{figure}[htpb]\centering
	\includegraphics[width=0.9\linewidth]{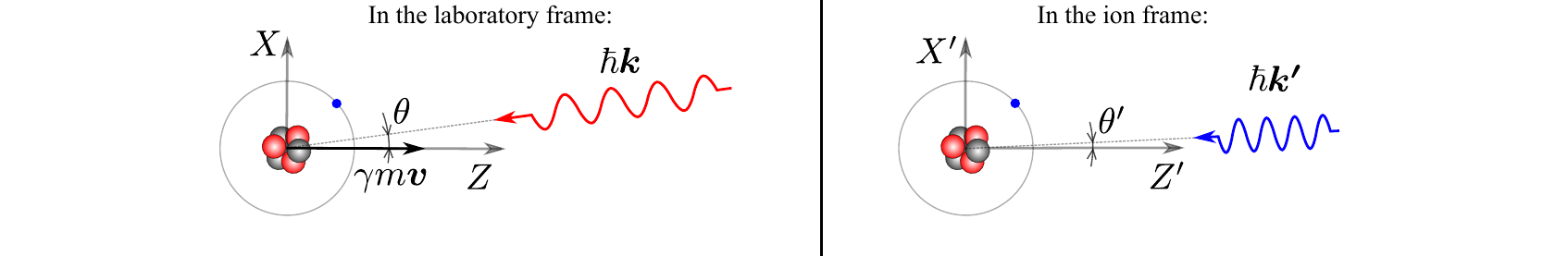}
	\caption{Photon absorption kinematics in the laboratory 
	and ion-rest frames.}
    \label{fig:Lorentz_transf_systems}
\end{figure}

To describe this process quantitatively, we use 4-vectors $(E/c, \boldsymbol{p})$. The Lorentz transformation of these 4-vectors,
for the  $z$-axis aligned with the direction of the ion motion,
can be written as 
\begin{equation}
\left( \begin{array}{c}E'/c \\ p_x'\\ p_y'\\ p_z' \end{array} \right) =
\begin{pmatrix}
\gamma _L & 0 & 0 & -\beta\gamma_L \\
0 & 1 & 0 &  0 \\
0 & 0 & 1 &  0 \\
-\beta\gamma_L & 0 & 0 &  \gamma_L \\
\end{pmatrix}
\left( \begin{array}{c}E/c \\ p_x \\ p_y \\ p_z \end{array} \right),
\label{eq:4vector}
\end{equation}
where the primed 4-vector components correspond to the ion-rest frame (Fig.~\ref{fig:Lorentz_transf_systems}). 

The ion 4-momentum is denoted, in the following,  by $(\gamma _L m c, \gamma _L m \boldsymbol{v})$ and the photon 4-momentum by $(\hbar \omega / c, \hbar \boldsymbol{k})$. 
Since the  photon momentum $\boldsymbol{p}$ is related  to the light's wave-vector $\boldsymbol{k}$ through $\boldsymbol{p}=\hbar \boldsymbol{k}$, where $\hbar$ is the reduced Planck constant, $k_x = -k\sin\theta$, $k_y = 0$, $k_z = -k\cos\theta$, and $k = \omega/c$, one can rewrite Eq.~\eqref{eq:4vector} as follows:
\begin{equation}
\left( \begin{array}{c} 1 \\ -\sin\theta' \\ 0 \\ -\cos\theta' \end{array} \right) \omega' =
\begin{pmatrix}
\gamma _L & 0 & 0 & -\beta\gamma _L \\
0 & 1 & 0 &  0 \\
0 & 0 & 1 &  0 \\
-\beta\gamma _L & 0 & 0 &  \gamma_L \\
\end{pmatrix}
\left( \begin{array}{c} 1 \\ -\sin\theta \\ 0 \\ -\cos\theta \end{array} \right) \omega.
\end{equation}
The relation between the photon frequency in the two frames is given by
\begin{equation}
    \omega' = ( 1 + \beta \cos\theta )\gamma _L \omega \approx \left( 1 + \beta - \beta\frac{\theta^2}{2} \right) \gamma _L \omega \approx 2 \gamma _L \omega.
    \label{eq:FrequencyFrameConversion}
\end{equation}
Eq.~\eqref{eq:FrequencyFrameConversion} shows that the frequency of the photon 
in the ion-rest frame is $2\gamma _L$-times larger than in the laboratory frame.
Such photons can excite transitions which are inaccessible 
for the  stationary ions.  

Tuning of the photon frequencies to the resonant atomic transitions
requires a precise control of the ion-beam angular 
divergence and its momentum spread. 
The angular divergence of the ion beam, $\Delta\theta$, may have an impact on the frequency (energy) spread of the photons in the ion rest frame.
However, since 
\begin{equation}
\omega' \sin\theta' = \omega \sin\theta,
\end{equation}
the spread
\begin{equation}
\Delta\theta' \approx \frac{\Delta\theta}{2\gamma _L},
\end{equation}
is significantly suppressed in the ion-rest frame,
as depicted in Fig.~\ref{fig:Lorentz_transf_systems}.
For example, the spread of the order of $\Delta\theta \approx 1\,$mrad
corresponds to the ion-frame frequency spread of only $\sim10^{-6}$. 
This shows that the angular divergence of the ion beam does
not contribute significantly to the effective photon-energy spread. 
A  significantly more important contribution comes from the energy
spread in the ion beam (typically $\Delta \gamma _L/ \gamma _L \approx 10^{-4}$). In order 
to achieve a high excitation rate of the ion beam,  the
frequency spread of the laser pulse should be tuned to be comparable to 
the ion-beam energy spread.

\begin{figure}[htpb]\centering
	\includegraphics[width=0.9\linewidth]{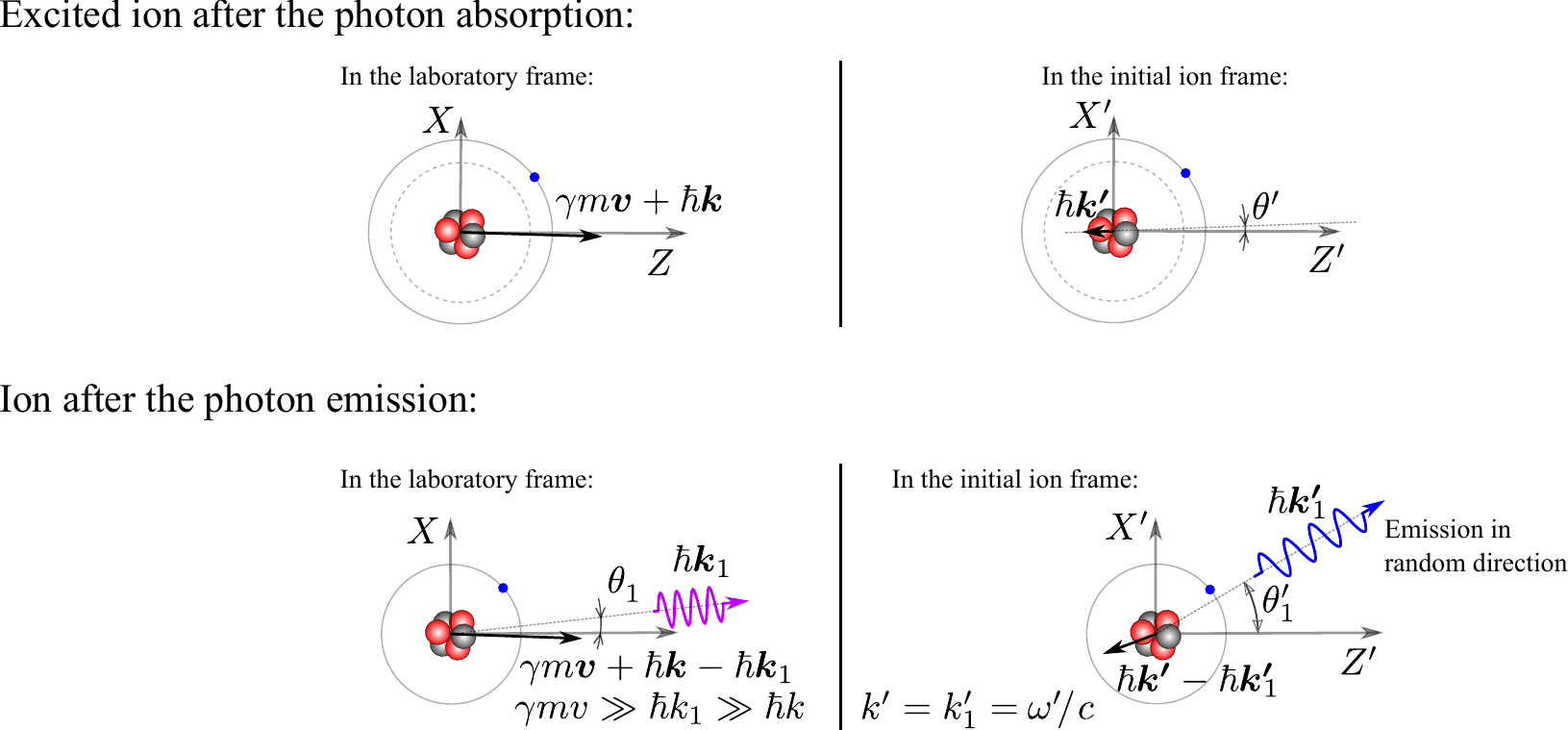}
	\caption{The excited ion after the photon absorption and the ion after the photon emission.}
    \label{fig:excited_ion_emitted_photon}
\end{figure}

The process of emission of a photon is  depicted in Fig.~\ref{fig:excited_ion_emitted_photon}, both in the laboratory 
and the ion-rest frame.
For  the atomic transitions considered in this paper as candidates 
for the beam cooling process, photons are emitted isotropically 
in the ion-rest frame. Their angular distribution is, however, 
strongly modified in the laboratory frame.
Let us consider the process where the photon is emitted in the same plane as the absorbed laser photon (the $x'$--$z'$ plane) at a random angle $\theta'_1$. 
In such a case, the photon 4-vector components are given by 
$k'_{1x} = k'\sin\theta'_1$ and $k'_{1z} = k'\cos\theta'_1$, 
and the inverse Lorentz transformation describes the relation between 
the emitted-photon parameters in the two frames:
\begin{equation}
\left( \begin{array}{c} 1 \\ ~\sin\theta_1 \\ 0 \\ \cos\theta_1 \end{array} \right) \omega_1 =
\begin{pmatrix}
\gamma _L & 0 & 0 & \beta\gamma _L \\
0 & 1 & 0 &  0 \\
0 & 0 & 1 &  0 \\
\beta\gamma _L & 0 & 0 &  \gamma _L \\
\end{pmatrix}
\left( \begin{array}{c} 1 \\ ~\sin\theta'_1 \\ 0 \\ \cos\theta'_1 \end{array} \right) \omega'.
\end{equation}
The emitted photon frequency is given by:
\begin{equation}
    \omega_1 = \gamma _L(1 +\beta\cos\theta'_1)\,\omega' 
    \approx 2\gamma_L^2(1 +\beta\cos\theta'_1)\, \omega.
    \label{eq:SecondaryPhotonEnergy}
\end{equation}
The laboratory-frame emission angle $\theta_1$ can be also calculated using the inverse Lorentz transformation:
\begin{equation}
\omega_1 \sin\theta_1 = \omega' \sin\theta'_1 ~\Rightarrow~ \sin\theta_1 = \frac{\sin\theta'_1} {\gamma _L(1 +\beta\cos\theta'_1)}.
\end{equation}
The resulting angular divergence of the emitted
photons in the laboratory frame is small for highly 
relativistic ions:  $\Delta\theta_1 \sim 1/\gamma _L $.

The change of the ratio of transverse to longitudinal momentum of the ion,  due to the photon emission,  is very small compared to the typical angular spread in the ion beam. Therefore, the main effect of the photon emission on the ion motion is the small loss of the ion total momentum.


\subsection{Cross section}

The cross section of the ion excitation by a photon with the frequency $\omega'$  \cite{Loudon2000,Metcalf1999} can be written as
\begin{equation}
\sigma = 2\pi^2 c r_e f_{12} g(\omega' - \omega'_0),
\label{cs}
\end{equation}
where $r_e$ is  classical electron radius, $f_{12}$ is the oscillator strength, $\omega'_0$ is the resonance frequency of the ion transition and $g(\omega'-\omega'_0)$ is the Lorentzian function: 
\begin{equation}
g(\omega' - \omega'_0) = \frac{1}{2\pi}\, \frac{\Gamma}{(\omega' - \omega'_0)^2 + \Gamma^2/4}\,,
\end{equation}
where $\Gamma$ is the atomic-resonance width related to  the lifetime of the excited ion $\tau'$:
\begin{equation}
\Gamma = \frac{1}{\tau'}\,.
\end{equation}
The atomic-resonance width can be expressed as
\begin{equation}
\Gamma = 2 r_e \omega_0'^2 f_{12}\frac{g_1}{cg_2},
\end{equation}
where $g_1, g_2$ are the degeneracy factors of the ground state and excited states, respectively.

The formula (\ref{cs}) can thus be rewritten in a simpler form:
\begin{equation}
\sigma(\omega' - \omega'_0) = \frac{\sigma_0}{1 + 4\tau'^2(\omega' - \omega'_0)^2},
\end{equation}
where
\begin{equation}
\sigma_0 = \frac{\lambda_0'^2 g_2}{2\pi g_1}, 
\end{equation}
 and $\lambda_0'=2\pi c/\omega_0'$ is the emitted photon  wavelength.

Equations presented in this Appendix are used in the simulations of the beam cooling process presented in  
Section~\ref{Ca-cooling}.

\end{appendices}

\end{document}